\documentclass[aps,pre,twocolumn,superscriptaddress,floatfix,nofootinbib]{revtex4}
\usepackage{amsmath,amssymb,amsfonts,bbm,graphicx,hyperref,times,color}
\usepackage{subfigure}
\usepackage{cleveref}

\newcommand{\ketbra}[2]{\vert #1 \rangle \langle #2 \vert}
\def\imm{\mathrm{i}}
\begin{document}
\title{Continuous-variable quantum probes for structured environments}
\author{Matteo Bina}
\email{matteo.bina@gmail.com}
\affiliation{Quantum Technology Lab, Dipartimento di Fisica, Universit\`a degli Studi di
Milano, I-20133 Milano, Italy}
\author{Federico Grasselli}
\affiliation{Institut f\"ur Theoretische Physik III, Heinrich-Heine-Universit\"at Düsseldorf, D-40225 D\"usseldorf, Germany}
\affiliation{Quantum Technology Lab, Dipartimento di Fisica, Universit\`a degli Studi di
Milano, I-20133 Milano, Italy}
\author{Matteo G. A. Paris}
\affiliation{Quantum Technology Lab, Dipartimento di Fisica, Universit\`a degli 
Studi di Milano, I-20133 Milano, Italy}
\affiliation{INFN, Sezione di Milano, I-20133 Milano, Italy}
\date{\today}
\begin{abstract}
We address parameter estimation for complex/structured systems and suggest
an effective estimation scheme based on continuous-variables quantum probes. 
In particular, we investigate the use of a single bosonic mode as a probe for 
Ohmic reservoirs, and obtain the ultimate quantum limits to the 
precise estimation of their cutoff frequency. We assume the probe prepared in 
a Gaussian state and determine the optimal working regime, i.e. the conditions 
for the maximization of the quantum Fisher information in terms of the initial 
preparation, the reservoir temperature and the interaction time. Upon investigating 
the Fisher information of feasible measurements we arrive at a remarkable simple 
result: homodyne detection of canonical variables allows one to achieve the ultimate 
quantum limit to precision under suitable, mild, conditions. Finally, upon exploiting 
a perturbative approach, we find the invariant {\em sweet spots} of the (tunable) 
characteristic frequency of the probe, able to drive the probe towards the optimal 
working regime. 
\end{abstract}
\maketitle
\section{Introduction}\label{s:introduction}
The dynamics of open quantum systems has been thoroughly investigated in recent years, 
due to its fundamental importance for decoherence and dissipation affecting 
quantum processes \cite{B.Petruccione,HarocheBook,Weiss,Caldeira}. 
The focus is on the reduced dynamics of a quantum system interacting with a surrounding environment, which can be modeled in a wide range of complexity, resulting in a master equation (ME) which rules the dynamics of the considered system and the observables associated to it. 
\par
The study of open quantum systems is multifaceted and presents multiple formulations 
depending on the features of the considered system, as well as those of the interaction
and the environment. The Lindblad form of the ME \cite{GKS,Lindblad} is suitable for 
describing Markovian (memory-less) dynamics and it is often employed in quantum optical 
systems \cite{Carmichael}. Whenever non-Markovian effects arise in the open dynamics, 
such as backflow of information and revival of entanglement \cite{BPL,VacchiniBackflow}, 
different approaches should be invoked to properly describe memory effects \cite{NonMarkov_Rev_Vega,BreuerVacchini}. The Brownian motion of a quantum system 
is a paradigmatic example for which analytical results are available \cite{B.Petruccione,StrunzQBM,HuPaz,Maniscalco2004,VasileParis}, and it will be our starting point to design 
effective characterisation schemes for structured environments. Indeed, previous 
results on stochastic fluctuating environments \cite{TrapaniBina,Trapani,RossiTrapani,Benedetti}, 
non-Markovian quantum jumps \cite{NMQJ}, stochastic master equations 
\cite{WisemanDiosi,GenoniMancini}, and  tunable non-Markovianity 
\cite{LiuNonMarkov,SmirneCialdi} have demonstrated the rich phenomenology of 
this scenario.
\par
The interaction between a quantum system and its harmonic environment, usually assumed 
in thermal equilibrium, is, in general, frequency-dependent and may be described in terms
of the {\em spectral density function} of the environment itself, which rules the range 
of frequencies accessible to the open system under investigation, and determines its 
rates of dissipation and decoherence \cite{Shnirman,Paavola,Legget,Martinazzo}. 
With the advent of state engineering \cite{Myatt,PiiloManiscalco}, it is possible to 
build a rich variety of structured environments and/or to simulate complex open 
quantum systems dynamics. This needs the full control of the parameters coming 
into play from both the Hamiltonian model and the dynamical ME for the reduced system.
\par
Often, the key parameters of a complex/structured environment are not directly observable,  
and a convenient strategy for their precise characterization is that of using the 
interaction with external probes. Open quantum systems may thus represent a resource 
for quantum characterization of structured environments, and valuable tools to optimise
the extraction of information are provided by quantum estimation theory (QET) 
\cite{Helstrom,ParisQET}. The basic concept is that from a given set of measured data, 
it is possible to build an optimal estimator and infer the value of the parameter of interest with a certain precision. The best attainable precision, provided a specific measurement, is limited by the  {\em Cramer-Rao bound} (CRB), whereas the {\em quantum Cramer-Rao bound} (QCRB) sets the best possible 
precision by optimizing over all the possible quantum measurements. Indeed, QET has been 
successfully employed in several fields as quantum phase transitions \cite{ZanardiPaunkovich,ParisZanardi,SalvatoriMandarino,AmelioBina,RossiBina}, characterisation 
of fluctuating  environments \cite{BenedettiBuscemi,ZwickKurizki,BenedettiParis}, quantum-optical 
interferometry \cite{Monras,GenoniOlivares,BinaSciRep}, quantum correlations \cite{BridaParis,GenoniBarbieri} and quantum thermometry \cite{BrunelliParis,CorreaAdesso}. 
\par
In this paper we address the problem of estimating, with the ultimate precision allowed by quantum mechanics, the cutoff frequency of Ohmic environments using continuous-variables quantum probes. In particular, we consider a single-mode quantum harmonic oscillator interacting with the structured environment, prepared in a generic Gaussian state. We discuss the optimal estimation strategy depending on the probe's initial parameters, such as thermal noise and the non-classical content provided by squeezing, and on the environment properties as temperature, interaction time and Ohmic spectral densities with an exponential cutoff. We will discuss the conditions to obtain the best possible precision in the estimation of the cutoff frequency, together with the analysis of a simple measurement scheme as the homodyne detection of the probe quadratures after the interaction with the environment.
\par
The paper is structured as follows. In Sec.~\ref{s:ME} we introduce the model describing the open dynamics of a quantum harmonic oscillator and its Gaussian solution, together with the tools of local QET to obtain the quantum Fisher information for Gaussian states. In Sec.~\ref{s:QE} we provide the results for the optimization of the quantum Fisher information as a function of the probe and the environment parameters, for a probe initialized in a squeezed-thermal Gaussian state. Moreover, we investigate the optimality conditions of the estimation strategy when a homodyne scheme is implemented. In Sec.~\ref{s:discussion} we extend the discussion to a generic single-mode Gaussian state, by introducing coherent energy in the initial probe state, whereas in Section \ref{s:ApprAlpha} we show analytic results in the weak coupling approximation, in order to discuss 
experimentally favourable conditions to obtain the optimal estimation of the parameter of interest. 
Sec.~\ref{s:conclusions} closes the paper with some concluding remarks.
\section{Master equation and Gaussian solution} \label{s:ME}
We begin by introducing the Hamiltonian of the model and the reduced dynamics of the probe, providing a general solution for a Gaussian-type initial preparation. The fundamental tools of local QET are shortly described and the recipe for computing the quantum Fisher information (QFI) of a generic single-mode Gaussian state is provided.
\subsection{The system-environment interaction}
The general description of a single quantum harmonic oscillator in interaction with a thermal environment is given by the following Hamiltonian:
\begin{equation}\label{Hgen}\begin{split}
H&=\frac{\hbar\omega_0}{2}\left( P^2 + X^2 \right)+ \sum_{n} \frac{\hbar\omega_n}{2} \left ( P_n^2+X_n^2 \right)+ \\
&-\alpha\, X\otimes \sum_{j} \hbar\gamma_j X_j,
\end{split} \end{equation}
where $X$ and $P$ are position and momentum operators, respectively, for the probe system, while $X_j$ and $P_j$ are position and momentum operators for the reservoir. We recall the generic quadrature operator expression: $X(\varphi)=(a\, {\rm e}^{-\imm\varphi}+a^\dag{\rm e}^{\imm\varphi})/\sqrt{2}$, where $X\equiv X(0)$ and $P\equiv X(\pi/2)$, with $a$ and $a^\dag$ are the annihilation and creation field operators, respectively. In Eq. (\ref{Hgen}) $\omega_j$ are the characteristic frequencies of the reservoir modes, $\gamma_j$ are the coupling frequencies between the system and the $j$-th reservoir mode, whereas $\alpha$ sets the dimensionless strength of the interaction between system and reservoir. From now on, without lack of generality, we set $\hbar = 1$. 
%
\par 
At $t=0$ we consider the global system in a factorized state $\varrho_0\otimes R$, where $\varrho_0$ is a generic initial state of the system oscillator and $R$ is the multi-mode equilibrium thermal state of the environment. In particular, the latter is the Gibbs state of the reservoir $R={\rm e}^{-\beta H_B}/\mathcal{Z}$, where $\beta=(k_B T)^{-1}$, with $k_B$ the Boltzmann constant and $T$ the equilibrium temperature of the reservoir, $H_B$ the free energy of the reservoir and $\mathcal{Z}$ the state normalization.
In order to describe the oscillator system dynamics, we refer to a very general master equation (ME) for the system oscillator evolved state $\varrho(t)$ in the weak-coupling regime, setting $\alpha\ll 1$:
 \begin{equation}\label{ME}\begin{split}
    \frac{d}{dt}\varrho(t)&=-\imm\big [ H_0,\varrho(t)\big ]+\imm\, r(t)\big[ X^2,\varrho(t)\big] \\ 
& -\imm\, \gamma(t)\big[X,\{ P,\varrho(t)\}\big ]  -\Delta(t)\big[X,\left[X,\varrho(t)\right]\big] \\
&+\Pi (t)\big[X,\left[P,\varrho(t) \right]\big] \,,
 \end{split} \end{equation}
 where $H_0=\omega_0\left( P^2 + X^2 \right)/2$ is the free Hamiltonian of the system and commutators and anti-commutators are represented, respectively, by $[\cdot\,,\cdot]$ and $\{\cdot\,,\cdot\}$.
This ME is local in time and can be obtained by means of a perturbative approach \cite{HuPaz}, or a time-convolutionless (TCL) method \cite{B.Petruccione}.
Besides the unitary contribution to the dynamics, given by $H_0$ in the first term of Eq.~(\ref{ME}), the second term is related to a time-dependent energy shift, the third one is a damping term and the last two are diffusion terms. The main feature of the time-local ME is that the time-dependent coefficients incorporate the non-Markovian behavior of the system dynamics.
In a perturbative approach, at second order in the coupling constant $\alpha$, these coefficients assume the following explicit expressions:
\setlength{\medmuskip}{0mu}
\begin{subequations}\label{CoefficientsME}\begin{align}
    r (\tau)&= \int_{0}^{\tau}d t' \cos(\omega_0 t')\int_{0}^{\infty}
      d\omega J(\omega) \sin(\omega t') \label{r_t} \\
    \gamma (\tau)&=\int_{0}^{\tau}dt' \sin(\omega_0
      t')\int_{0}^{\infty}  d\omega J(\omega) \sin(\omega t') \label{gamma_t}\\
         \Delta (\tau)&=
     \int_{0}^{\tau}dt' \int_{0}^{\infty}
        d\omega \big [1+2N(\omega)\big ] J(\omega) \cos(\omega_0 t')\cos(\omega t') 
        \label{delta_t} \\
    \Pi (\tau)&=
        \int_{0}^{\tau}dt' \sin(\omega_0 t') \int_{0}^{\infty}
        d\omega  [1+2N(\omega)\big ] J(\omega) \cos(\omega t') \,,
        \label{pi_t}
\end{align} \end{subequations}
where $T$ is the reservoir temperature, $J(\omega)=\alpha^2\sum_j \frac{\gamma_j}{2}\delta(\omega-\omega_j)$ is a generic spectral density characterizing the structured environment, and $N(\omega)=\big ({\rm e}^{\beta\omega}-1\big )^{-1}$ is the mean thermal excitation number of the reservoir.
Following the superoperator formalism introduced by Intravaia {\it et al.} in Ref. \cite{IntraManMess}, which relies only on the weak-coupling approximation and is independent on the type of reservoir, we can write the general solution of the ME (\ref{ME}) in terms of the characteristic function in cartesian coordinates $\vec{z}=(x\, ,\, p)$:
\begin{equation}
  \chi [\vec{z}\,](t)=
   {\rm e}^{-\vec{z}^{\,T}\overline{W}(t)\vec{z}}
      \chi\left[e^{-\frac{\Gamma (t)}{2}}R^{-1}(t)\vec{z}\, \right] (0) \, ,
      \label{QCFt}
\end{equation}
where 
\begin{subequations}\begin{align}
 \Gamma (t) &= 2\int_{0}^{t} \gamma(\tau) d\tau  \label{Gamma} \\
  R(t) &\approx
   \left( \begin{array}{cc}
      \cos(\omega_0 t) & \sin(\omega_0 t) \\ [1.5ex] -\sin(\omega_0 t) & 
       \cos(\omega_0 t)
    \end{array}
   \right)  \label{R(t)}  \\
 \overline{W}(t)&={\rm e}^{-\Gamma(t)}\big [ R^{-1}(t) \big ]^T W(t) R^{-1}(t) \label{Wbar}\\
 W(t)&=\int_0^t{\rm e}^{\Gamma(\tau)}R^T(\tau)M(\tau)R(\tau)d\tau \\
  M(\tau)&=\left( \begin{array}{cc}
    \Delta(\tau) &-\frac{\Pi (\tau)}{2} \\ [1.5ex] -\frac{\Pi (\tau)}{2} & 
       0
    \end{array}
   \right) .
\end{align}\end{subequations}
By noticing that Hamiltonian (\ref{Hgen}) is at most bilinear in the bosonic field modes, it induces a Gaussian evolution map, thus preserving the Gaussian character of initial Gaussian states. For a start, we focus on a particular class of single-mode Gaussian states, the squeezed thermal states (STS), which can be written in the form $\varrho_0=S(\xi)\nu(n_{\text{th}})S^\dag(\xi)$ (we will consider a more general Gaussian state later on). The interplay between pure non-classical character, induced by the squeezing operator $S(\xi)=\exp\{(\xi a^2-\xi^*a^{\dag\;2})/2\}$, with $\xi=|\xi| {\rm e}^{\imm \theta}$ the complex squeezing parameter, and classical noise given by the mean number of thermal excitations $n_\text{th}$ (state $\nu(n_{\text{th}})$ is a single-mode thermal state), will set specific conditions for the initial preparation of the probe state $\varrho_0$ for the optimal estimation strategy, as discussed in the next sections. The Wigner function of a single-mode Gaussian state $\varrho$ is of the form
\begin{equation} \label{WGaussian}
\mathcal{W}[\varrho](\vec{z})=\frac{\exp \big [-\frac12 (\vec{z}-\vec{\delta})^T\sigma^{-1}(\vec{z}-\vec{\delta})\big ]}{2\pi \sqrt{\text{det}\,\sigma}} \, .
\end{equation}
where $\sigma$ and $\vec{\delta}$ are, respectively, the covariance matrix (CM) and the first-moment vector of the Gaussian state \cite{FerraroOlivaresParis,Braustein,Olivares,AdessoReview}. In our case, the initial STS of the probe has a null first-moment vector $\vec{\delta}_0=0$, whereas the CM is $\sigma_0=\Sigma_{\xi,\theta} \sigma_\text{th} \Sigma_{\xi,\theta}^T$, with
\begin{subequations}\begin{align}
\sigma_\text{th}&=\left(\frac{1}{2}+n_\text{th}\right) \mathbb{I} \\
\Sigma_{\xi,\theta}&= \cosh\xi\, \mathbb{I}+\sinh\xi
 \left(
 \begin{array}{cc}
  \cos\theta & \sin\theta \\ -\sin\theta & \cos\theta
 \end{array}
 \right)  \, . \label{covmatrix}
\end{align} \end{subequations}
The probe state evolved under the ME (\ref{ME}), is a Gaussian single-mode state with null first-moment vector $ \vec{\delta} = 0$ and CM of the form
 \begin{equation}
   \sigma= 2\overline{W}(t)+
    e^{-\Gamma (t)} R(t)
  \sigma_0 R^{-1}(t) \label{CM_t}  \,.
 \end{equation}
We underline that the evolved Gaussian state has been obtained under very general conditions, by performing only the weak-coupling approximation for $\alpha\ll 1$, without implementing secular and Markov approximations. Being a ME local in time, the memory effects are contained in the time-dependent coefficients (\ref{CoefficientsME}), whereas Markovian and non-Markovian regimes depend on the evolution timescales of system and reservoir, $t_S$ and $t_R$ respectively. As it will be discussed, the non-Markovian character does not bring any advantage in optimizing the parameter estimation, thus, in presenting the main results, we will consider both secular and Markov approximations.

%

\subsection{Local QET for Gaussian states}
The main goal of this paper is to obtain the optimal measurement scheme for estimating an unknown parameter $\lambda$ of a physical system, not directly associated to a measurable observable. Local QET provides the tools to calculate the ultimate limits, imposed by quantum mechanics, to the precision of the estimation protocol, in terms of the variance of any estimator of the parameter $\lambda$. This unavoidable uncertainty has a lower limit, namely the quantum Cram\'er-Rao bound (QCRB), which is related to the quantum Fisher information (QFI) $H_\lambda$
\begin{equation}\label{QCRB}
\text{Var}_\lambda\geq\frac{1}{M H(\lambda)}\, ,
\end{equation}
where $M$ is the statistical scaling factor given by the $M$ outcomes associated to the optimal quantum measurement $\mathcal{L}_\lambda$, which defines the QFI through 
\begin{equation}\label{QFI}
H(\lambda)\equiv\mathrm{Tr}[\varrho_\lambda\mathcal{L}^2_\lambda].
\end{equation}
The selfadjoint operator $\mathcal{L}_\lambda$ is called symmetric logarithmic derivative (SLD) and it is implicitly defined by $\partial_\lambda\varrho_\lambda \equiv (\mathcal{L}_\lambda\varrho_\lambda+\varrho_\lambda\mathcal{L}_\lambda)/2$. In general, Eq. (\ref{QCRB}) is independent on the measurement scheme and the corresponding sets of outcomes, but depends only on the system quantum state. Considering, instead, a particular measurement, given by an observable $\mathcal{O}$ with outcomes $\{x\}=\{x_1,\ldots,x_M\}$, the standard Cram\'er-Rao bound (CRB) in classical statistical theory, provides a bound to the precision of an unbiased estimator of the parameter $\lambda$ given by
\begin{equation}\label{CRB}
\text{Var}_\lambda\geq\frac{1}{M F(\lambda)}\, ,
\end{equation}
where  $F(\lambda)$ is the Fisher information (FI) associated to the observable $\mathcal{O}$. The FI now depends on the statistics $p(x|\lambda)$ of the particular measurement scheme, which is parameter-dependent, and reads
\begin{equation}\label{FI}
F(\lambda)  = \int d\, x \,  \frac{\big [ \partial_\lambda p(x | \lambda)\big ]^2}{p(x | \lambda)} \, .
\end{equation}
A straightforward calculation \cite{ParisQET} brings to the inequality $H(\lambda)\geq F(\lambda)$ and the optimal measurement scheme is the one for which this inequality is saturated, i.e. the CRB saturates the QCRB. In order to be precise, we note that the last statement is valid for quantum parameter-independent measurements, otherwise one should refer to more general bounds on the FI \cite{SevesoRossiParis}. On top of this, another significant quantity in assessing the performances of an estimator, is the signal-to-noise ratio (SNR), which is defined as $\text{SNR}\equiv \lambda^2/\text{Var}_\lambda$. Using the QCRB (\ref{QCRB}) and CRB (\ref{CRB}), it is natural to define the corresponding best possible SNRs for, respectively, the QFI and the FI:
\begin{equation}\label{SNRFH}
\text{SNR}_H\equiv \lambda^2 H(\lambda) \qquad \text{SNR}_F\equiv \lambda^2 F(\lambda) \, .
\end{equation}
\par 
In the context of single-mode Gaussian states, represented by the Wigner function (\ref{WGaussian}), it is possible to derive an explicit expression for the QFI. The SLD of a Gaussian state can always be written as a degree-2 polynomial of the position and momentum operators $\vec{Z}=(X\,,\,P)$:
 \begin{equation}
  \mathcal{L}_\lambda=\vec{Z}^{\,T}\Phi\vec{Z} + \vec{Z}^{\,T} \vec{\zeta} - \nu
   \label{SLDgaussian}
 \end{equation}
where $\vec{\zeta}=\Omega^T\sigma^{-1}\langle \dot{\vec{Z}}\rangle$ (with $\dot{f}\equiv\partial_\lambda f$), $\nu=\mathrm{Tr}[\Omega^T \sigma \,\Omega\, \Phi]$,
\begin{equation}\label{Omega}
\Omega= \left(
 \begin{array}{cc}
  0 & 1 \\ -1 & 0
 \end{array}
 \right) \, , 
\end{equation}
 and $\Phi$ is a symmetric real 2x2 matrix, implicitly defined by
\begin{equation}\label{PhiL}
 \dot{\sigma} = 2 \sigma\,\Phi\,\sigma - \frac{1}{2}\Omega\,\Phi\,\Omega^T \, .
\end{equation}
In order to obtain an explicit and computable expression of the QFI (\ref{QFI}), we perform a symplectic diagonalization of the CM, namely $\Sigma\sigma \Sigma^{T}=\sigma_\Sigma=\text{diag}(d,d)$, where $d$ is the symplectic eigenvalue and $\Sigma$ is a proper symplectic matrix satisfying $\Sigma\,\Omega\, \Sigma^T=\Omega$. By applying this symplectic transformation to Eq. (\ref{PhiL}) and combining the result with Eqs.~(\ref{QFI},\ref{SLDgaussian}), we obtain an explicit expression for the QFI of a single-mode Gaussian state as a function of the parameter of interest $\lambda$:
\begin{equation}
    H(\lambda)=\frac{1}{2{d}^4-1/8}\left\lbrace
     {d}^4\mathrm{Tr}[(\sigma^{-1}\dot{\sigma})^2]
      -\frac{1}{4}\mathrm{Tr}[(\Omega\dot{\sigma})^2]
      \right\rbrace  \label{QFIfinal} \, ,
\end{equation}
where we assumed a null first-moment vector, in accordance with the choice of the initial STS of the probe system. We followed the approach adopted by Jiang in Ref.~\cite{Jiang}, but we point out that the same result can be obtained employing the method illustrated by Pinel \emph{et al.} in Ref.~\cite{Pinel}.
\par
The explicit expression of the QFI (\ref{QFIfinal}) depends on the parameters of the system and the reservoir, appearing in the Hamiltonian (\ref{Hgen}) and in the coefficients (\ref{CoefficientsME}) of the ME. With this expression, though quite complicated, it is possible to evaluate the best precision to the estimation of the parameter of interest allowed by the QCRB for every choice of system-reservoir setting and for every dynamical regime. Instead, in order to find the conditions for the optimization of the QFI, in the next Section we will resort to approximated analytical expressions which grab the physics behind the considered interaction between a continuous-variable probe and a structured environment.

\section{Quantum estimation of the cutoff frequency} \label{s:QE}
\begin{figure}[t!]
\center
\includegraphics[width=0.48\textwidth]{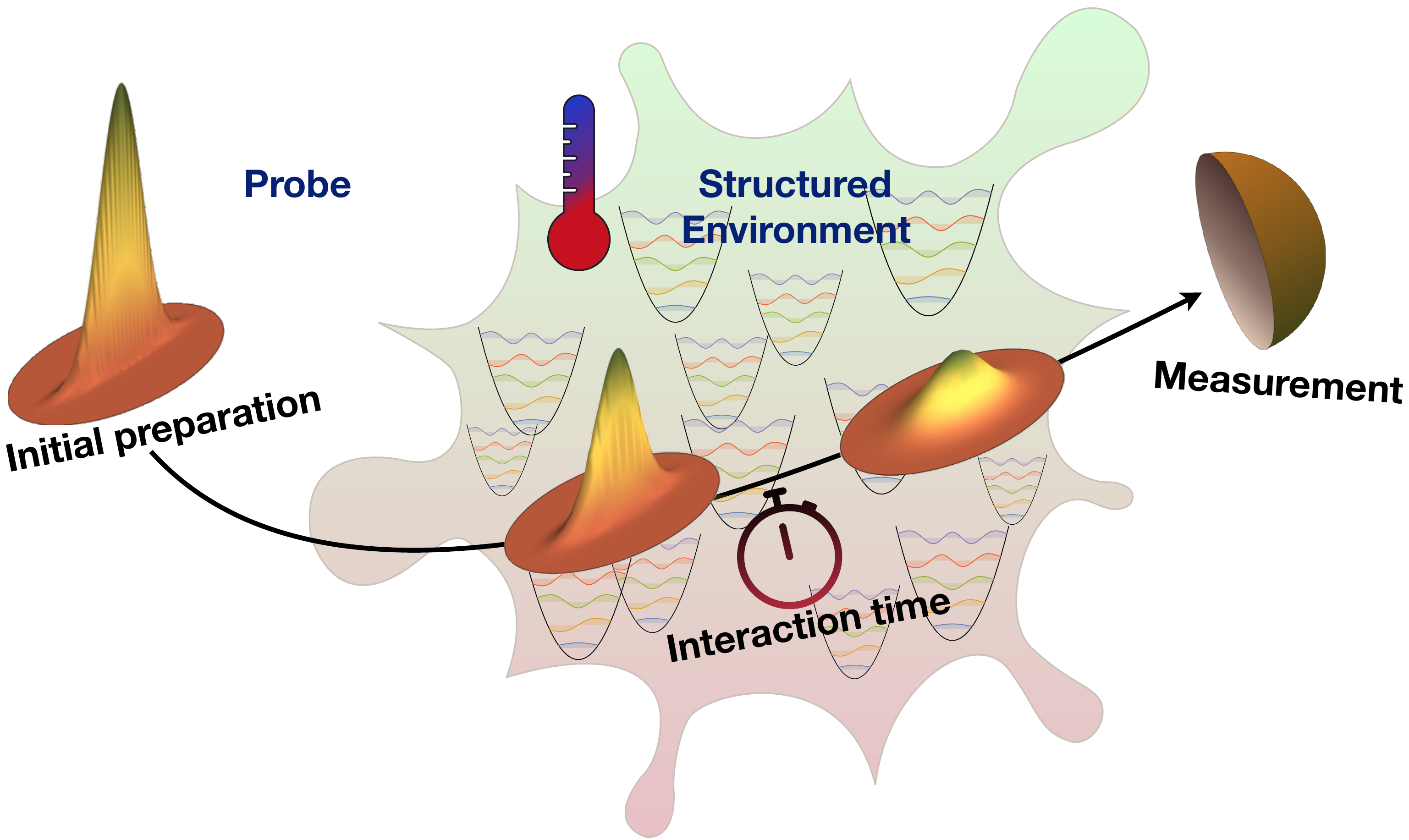}
\caption{(Color online) Scheme for the probing of a structured thermic environment with a single-mode probe prepared in a generic Gaussian state.} \label{f:schema}
\end{figure}
In this Section we will follow the estimation protocol for the quantum probing of a structured environment, pictorially described in Fig.~\ref{f:schema}. Firstly, we show the results regarding the QFI as a function of the parameters involved in the interaction with the reservoir. Then, we optimize the QFI with respect to the initial probe state parameters. Eventually, the measurement stage is optimized when a homodyne scheme is employed on the evolved state of the probe.

\subsection{QFI and optimization of the probe}
Even though the solution of the ME (\ref{ME}) for an initial Gaussian state is very general and, thus, we are able to provide results for generic values of the involved parameters, we will focus on some approximated regimes, which are the ones containing the most important results concerning the optimization of the parameter estimation protocol.
\par
The first reasonable approximation to perform is the secular approximation, which allows to neglect the fast oscillating terms in Eq. (\ref{Wbar}), by choosing a coarse-grained dynamical regime $\omega_0 t\gg 1$  \cite{VasileParis,Paavola}. The secularly approximated CM of the evolved probe state reads
\begin{equation}\label{CM_Secular}
 \sigma= \Delta_\Gamma (t)\, \mathbb{I} +e^{-\Gamma (t)} R(t)\sigma_0 R^{-1}(t)\, ,
\end{equation}
where we defined
\begin{equation}\label{DeltaGamma}
\Delta_\Gamma (t) \equiv e^{-\Gamma (t)}\int_{0}^{t}e^{\Gamma (\tau)}\Delta (\tau) d\tau \,.
\end{equation}
Implementing these expressions into Eq.~(\ref{QFIfinal}), we obtain the QFI for a generic initial STS under secular approximation, which is too heavy to report here. As we are interested in probing an unknown parameter of the structured environment, we point out that the dependence on the parameter of interest $\lambda$ is carried implicitly by the generic spectral density $J_\lambda(\omega)$ and, thus, by the coefficients $\Gamma$ and $\Delta_\Gamma$.
\par
As a further consideration, we consider evolution times $t$ much greater than the characteristic correlation time of the structured reservoir $\tau_R$. This allows us to perform the Markov approximation, enabling to extend to infinity the interval of integration in Eqs.~(\ref{CoefficientsME}), namely $\tau\to\infty$ \cite{Carmichael}. As it will become soon clearer, the QFI simply increases with the interaction time (up to large time intervals), so for our purposes the non-Markovian regime does not bring any effect of enhancement of the QFI, thus supporting the Markov approximation. In the Markov regime, we obtain the limiting constant values of the coefficients $\gamma(\tau)$ and $\Delta(\tau)$, the only ones ruling the system dynamics under the secular approximation:
\begin{subequations}\begin{align}
\gamma_M&=\lim_{\tau\to\infty}\gamma(\tau)=\frac{\pi}{2}J_\lambda(\omega_0) \\
\Delta_M&=\lim_{\tau\to\infty}\Delta(\tau)=\pi \coth\left(\frac{\omega}{2k_B T}\right) J_\lambda(\omega_0) \, .
\end{align}\end{subequations}
The CM (\ref{CM_Secular}) now has a simple time dependence in terms of the Markovian coefficients
\begin{subequations}\label{Markov}\begin{align}
\Gamma_M (t)&= \pi J_\lambda(\omega_0) t \\
\Delta_{\Gamma_M}(t)&= \frac{1}{2}\coth\left(\frac{ \omega_0}{2 k_B T}\right) \left(1-e^{-\pi J_\lambda(\omega_0) t}\right) \, .
\end{align}\end{subequations}
%
%
%
\begin{figure}[t!]
\center
\includegraphics[width=0.237\textwidth]{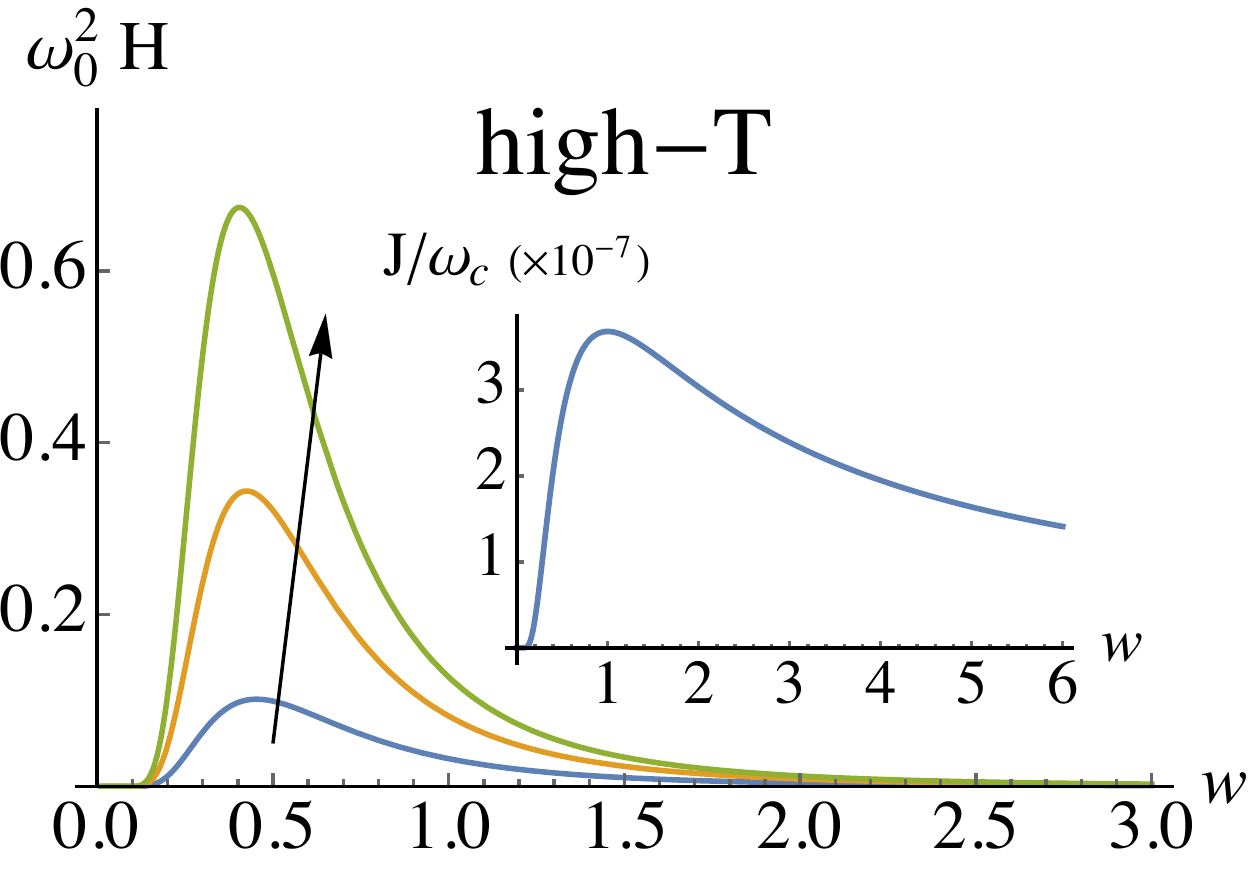} \includegraphics[width=0.237\textwidth]{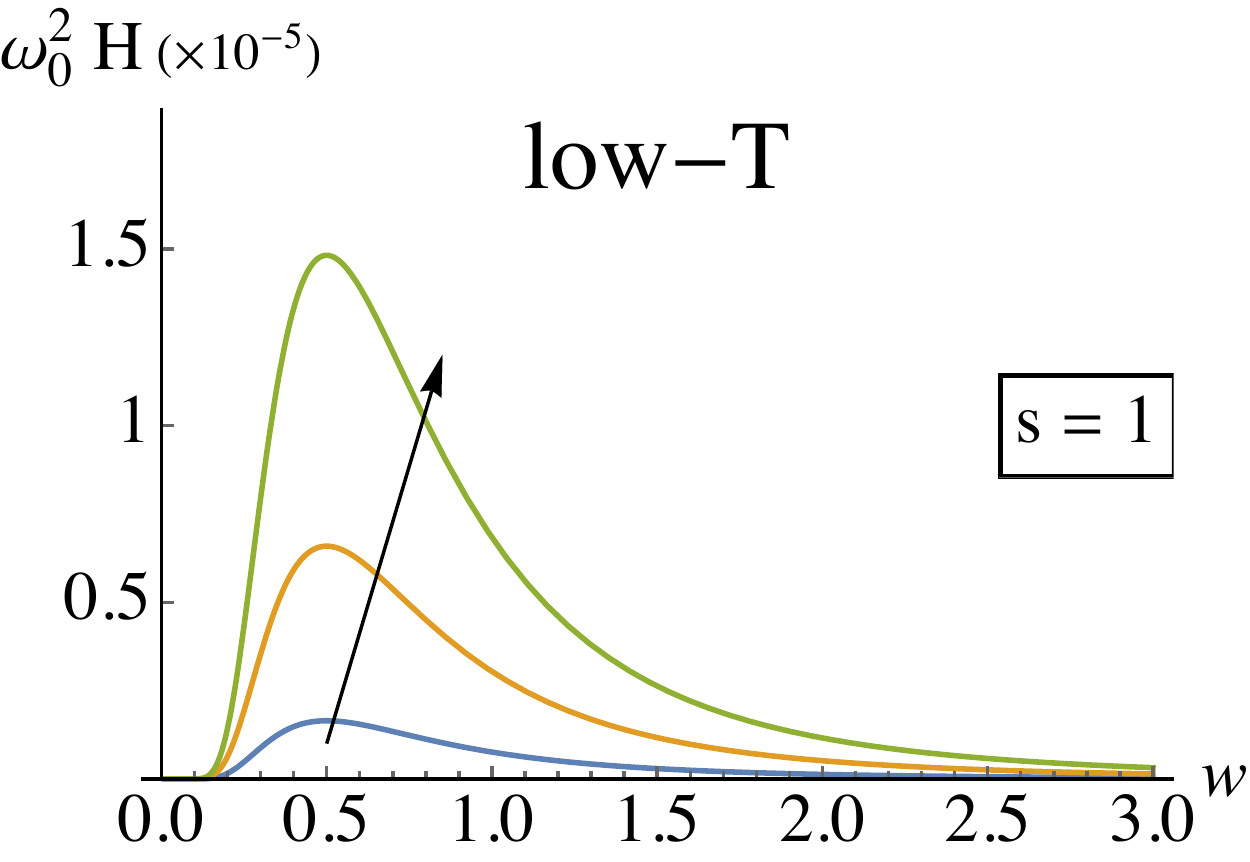}\\
\includegraphics[width=0.237\textwidth]{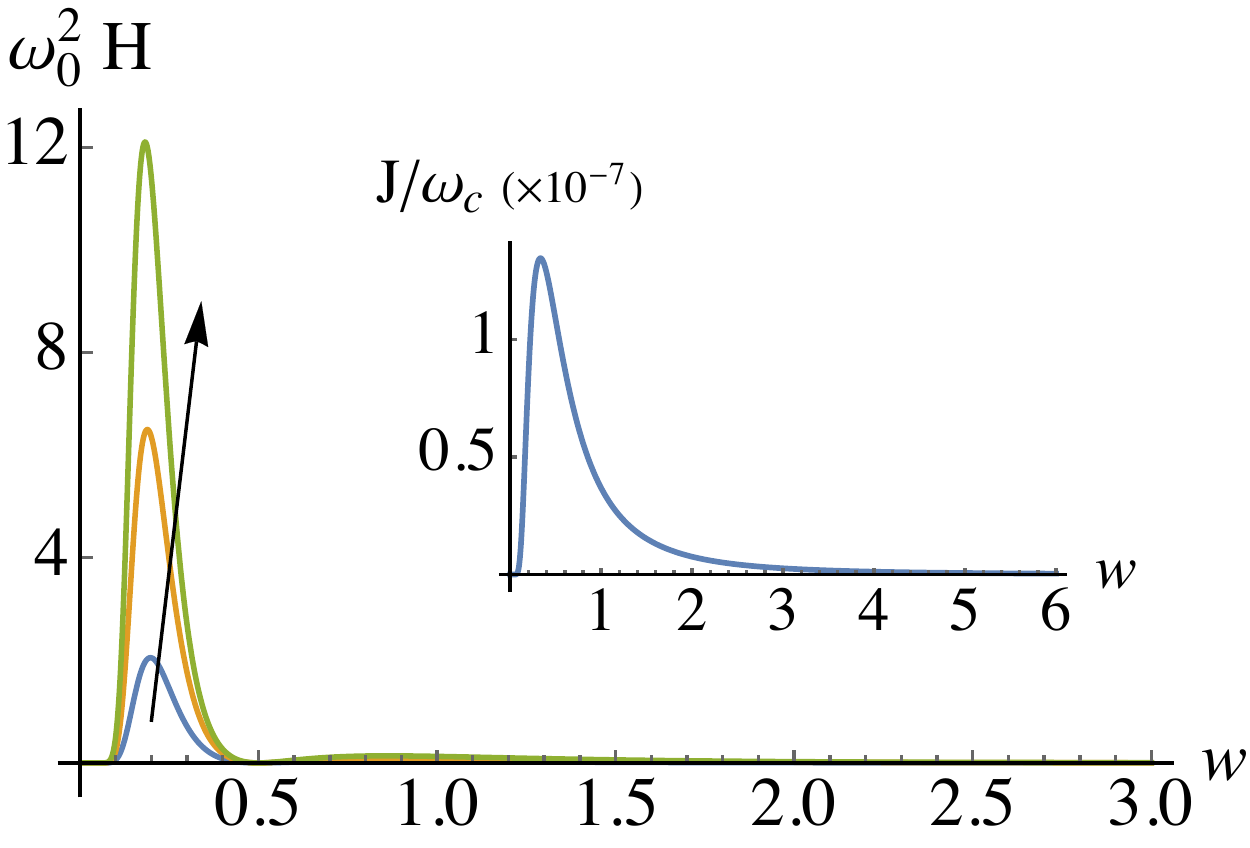} \includegraphics[width=0.237\textwidth]{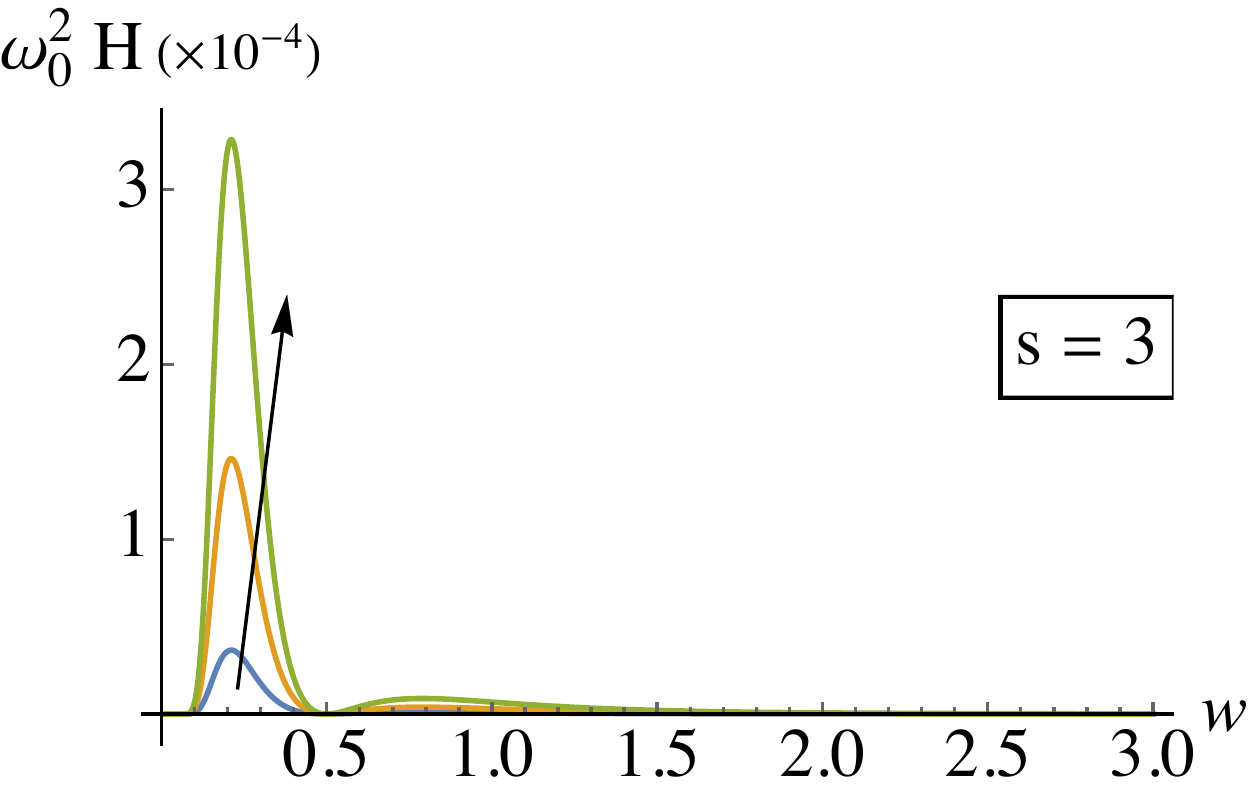}\\
\includegraphics[width=0.237\textwidth]{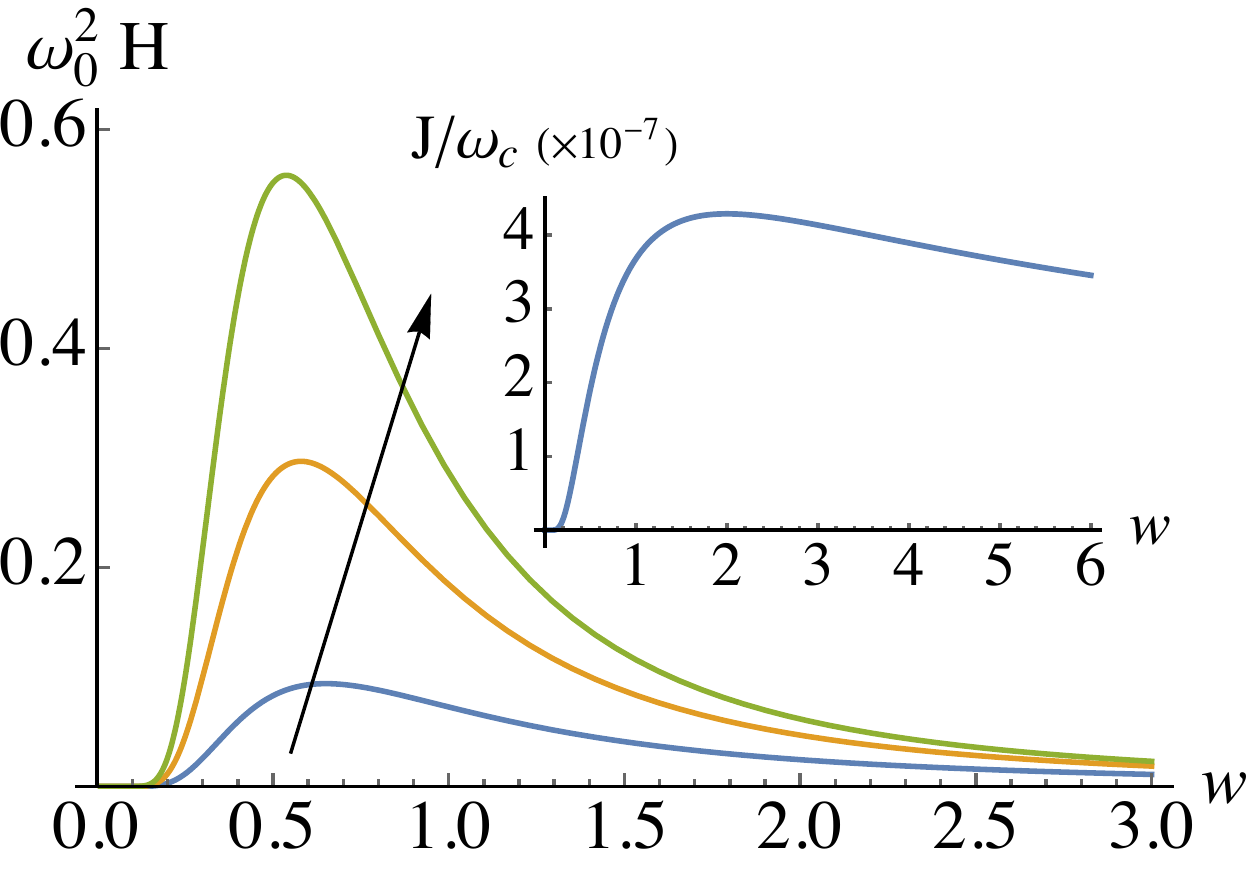} \includegraphics[width=0.237\textwidth]{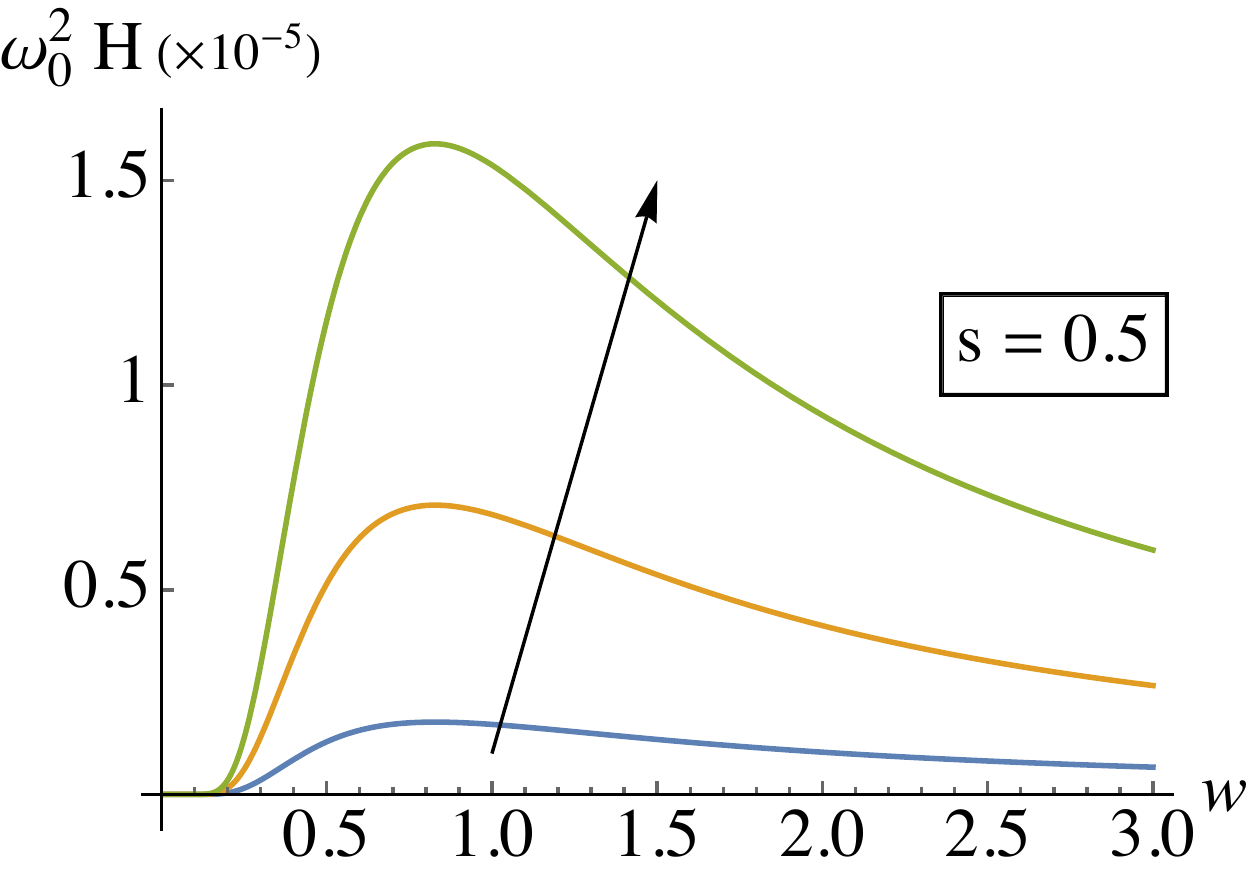}
\caption{(Color online) Plots of the QFI, rescaled by $\omega_0^2$, as a function of $\text{\it{w}}=\omega_c/\omega_0$ for two temperature regimes (columns), high-T with $k_B T=10^2\,\omega_0$ and low-T with $k_B T=10^{-2}\,\omega_0$, and for three cases of structured environments (rows), with $s=1$, $s=3$ and $s=0.5$. In each plot we show three curves corresponding to different interaction times, as indicated by the arrows: $\omega_0 t = 10^3$, $\omega_0 t = 2\cdot 10^3$ and  $\omega_0 t = 3\cdot 10^3$. In the insets we show the spectral density rescaled by $\omega_c$. The other parameters are set as $\xi=1$, $n_\text{th}=2$ and $\alpha=10^{-3}$.} \label{f:QFI_wc_s_time}
\end{figure}
So far, our considerations are valid for any type of structured environments with a generic spectral density $J_\lambda(\omega)$. In order to provide some examples and obtain quantitative results, we consider the following families of spectral density in the continuum-mode limit for the reservoir:
\begin{equation}
   J_{\omega_c}(\omega)=\alpha^2\,\omega_c {\left(\frac{\omega}{\omega_c}\right)}^s {\rm e}^{-\omega/\omega_c} \,,
    \label{J(w)}
\end{equation}
which describe the Ohmic ($s=1$), sub-Ohmic ($s<1$) and super-Ohmic ($s>1$) reservoirs \cite{Legget,Shnirman}. The exponential cutoff is ruled by the frequency $\omega_c$, which limits the accessibility to the environmental frequencies by the probe system (see the insets of Fig.~\ref{f:QFI_wc_s_time}). In the following our parameter of interest will be $\lambda\to\omega_c$.
\par
The cutoff frequency $\omega_c$ represents the reservoir parameter to be estimated and we will apply the QET tools outlined above to provide the ultimate bounds to the precision in its estimation. Firstly, we study the behavior of the QFI $H(\omega_c)$ as a function of the reservoir parameters, fixing a generic initial STS of the probe. In Fig.~\ref{f:QFI_wc_s_time} we show the QFI $H(\omega_c)$ as a function of the ratio $\text{\it{w}}\equiv\omega_c/\omega_0$ in two different regimes of reservoir temperature, the high- and low-temperature regimes with $k_B T=10^2\,\omega_0$ and $k_B T=10^{-2}\,\omega_0$, respectively, for three characteristic reservoir parameters $s=1$, $s=3$ and $s=0.5$. Each plot contains three curves corresponding to different and increasing interaction times (from bottom to top), specifically $\omega_0 t = 10^3$, $\omega_0 t = 2\cdot 10^3$ and  $\omega_0 t = 3\cdot 10^3$, in the respect of Markov approximation $t\gg \tau_R\equiv 1/\omega_c$. The QFI $H(\omega)$ is higher in the high-temperature case than in the low-temperature case and it increases for longer interaction times. This last aspect, is valid up to a certain interaction time, after which the QFI decreases, as one may expect from the fact that there exists a stationary state of the probe due to a thermalization process with the environment.
\par
\begin{figure}[t!]
\center
\includegraphics[width=0.235\textwidth]{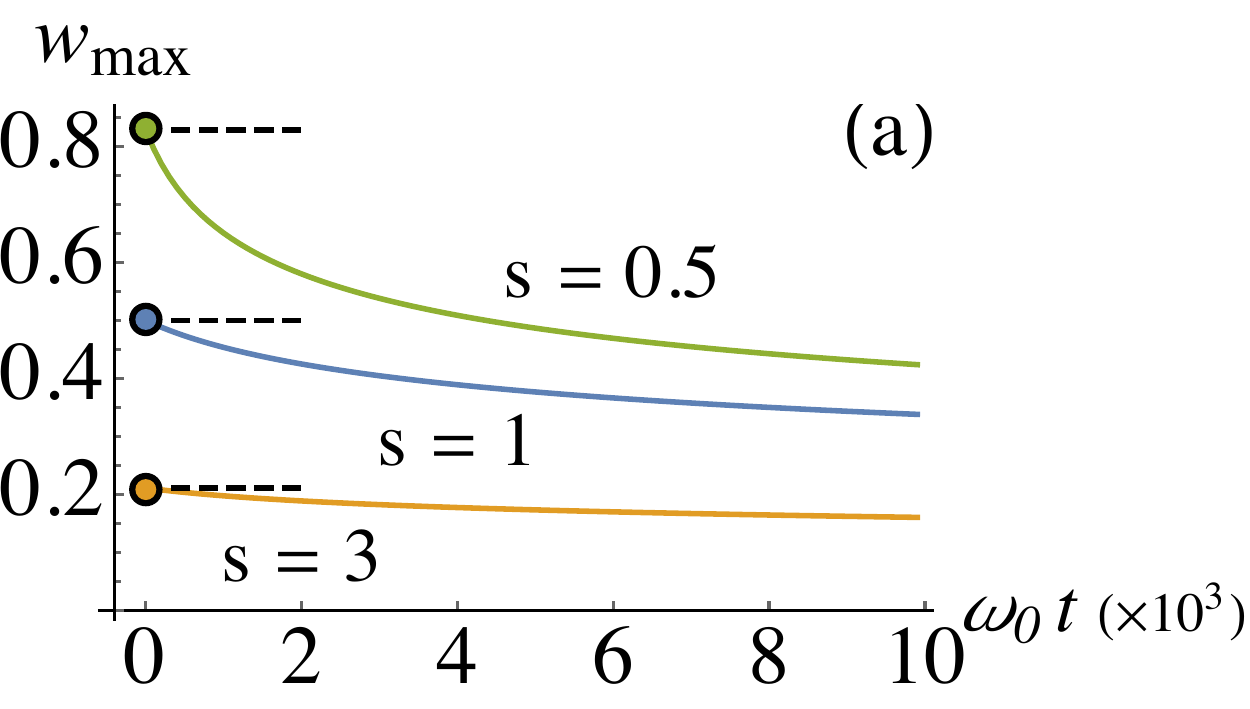}\hspace{1mm}
\includegraphics[width=0.235\textwidth]{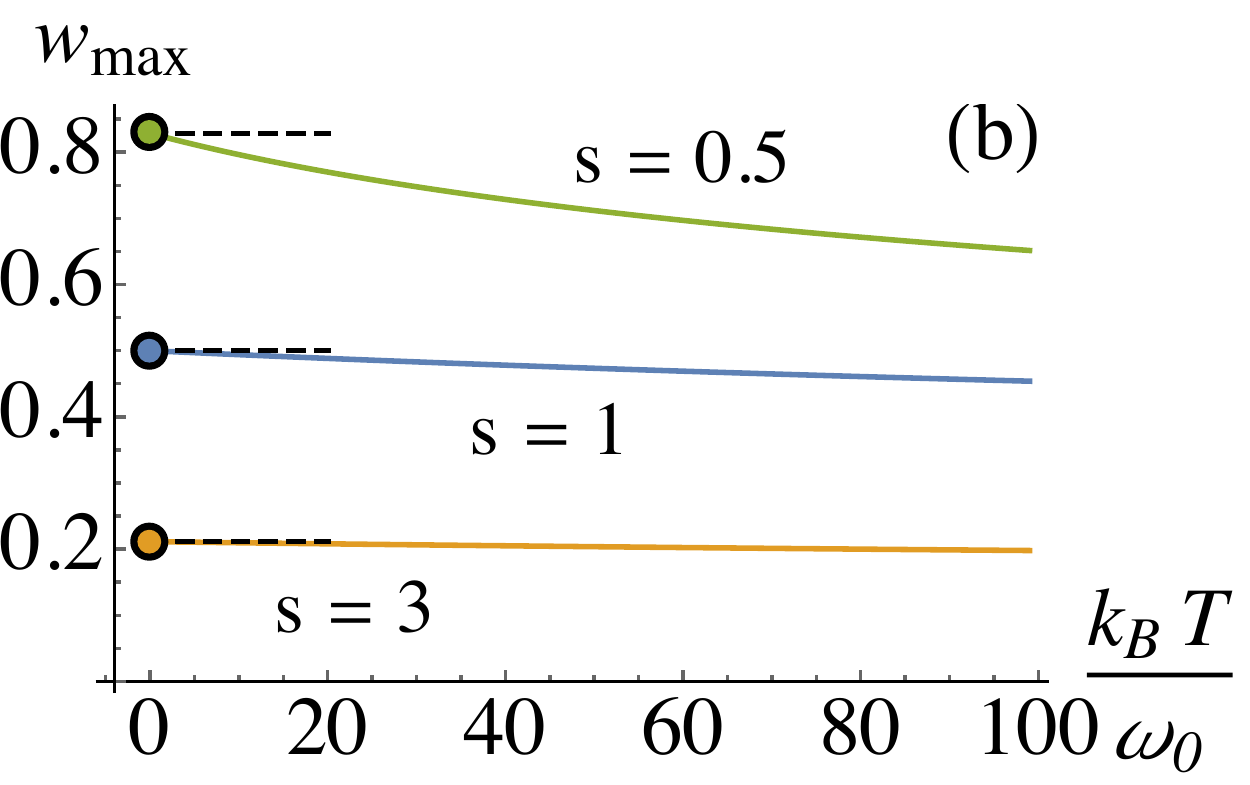}\\
\includegraphics[width=0.235\textwidth]{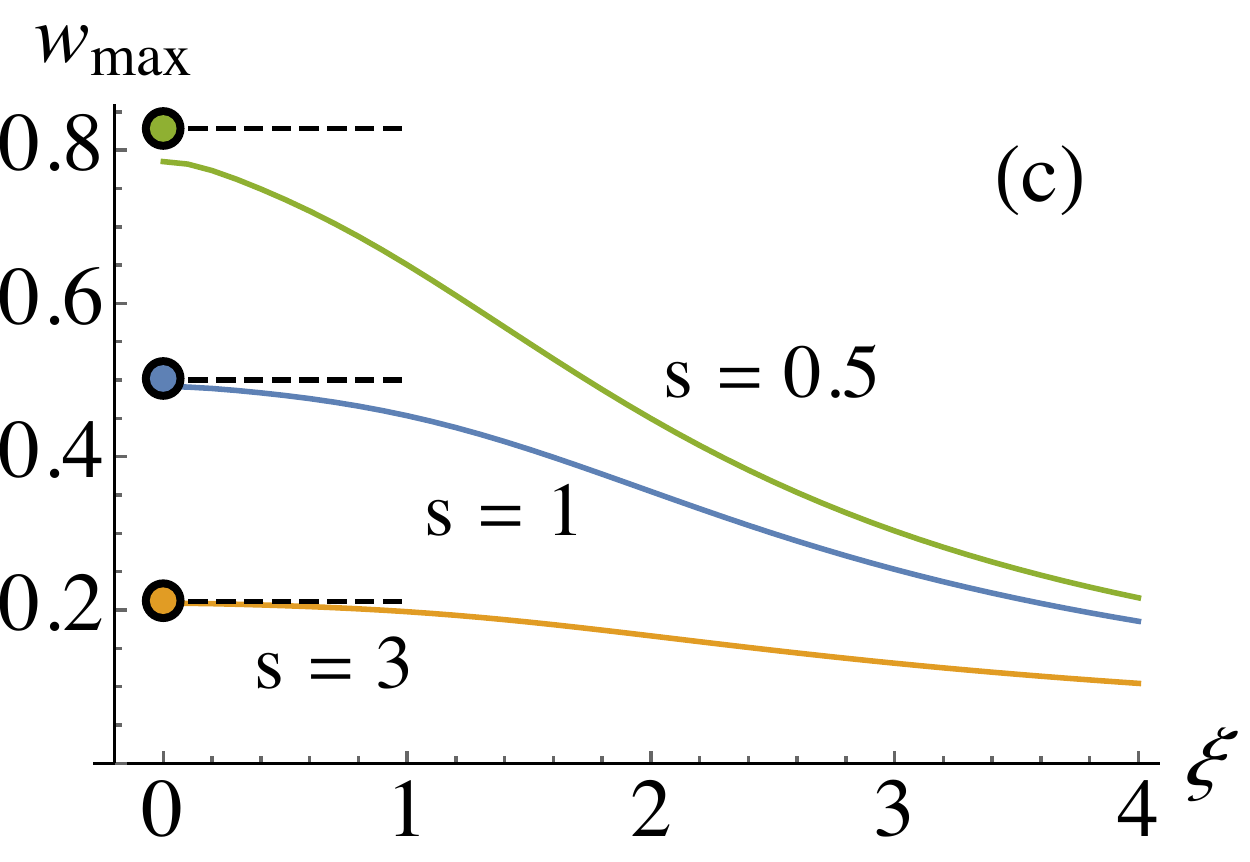}\hspace{1mm}
\includegraphics[width=0.235\textwidth]{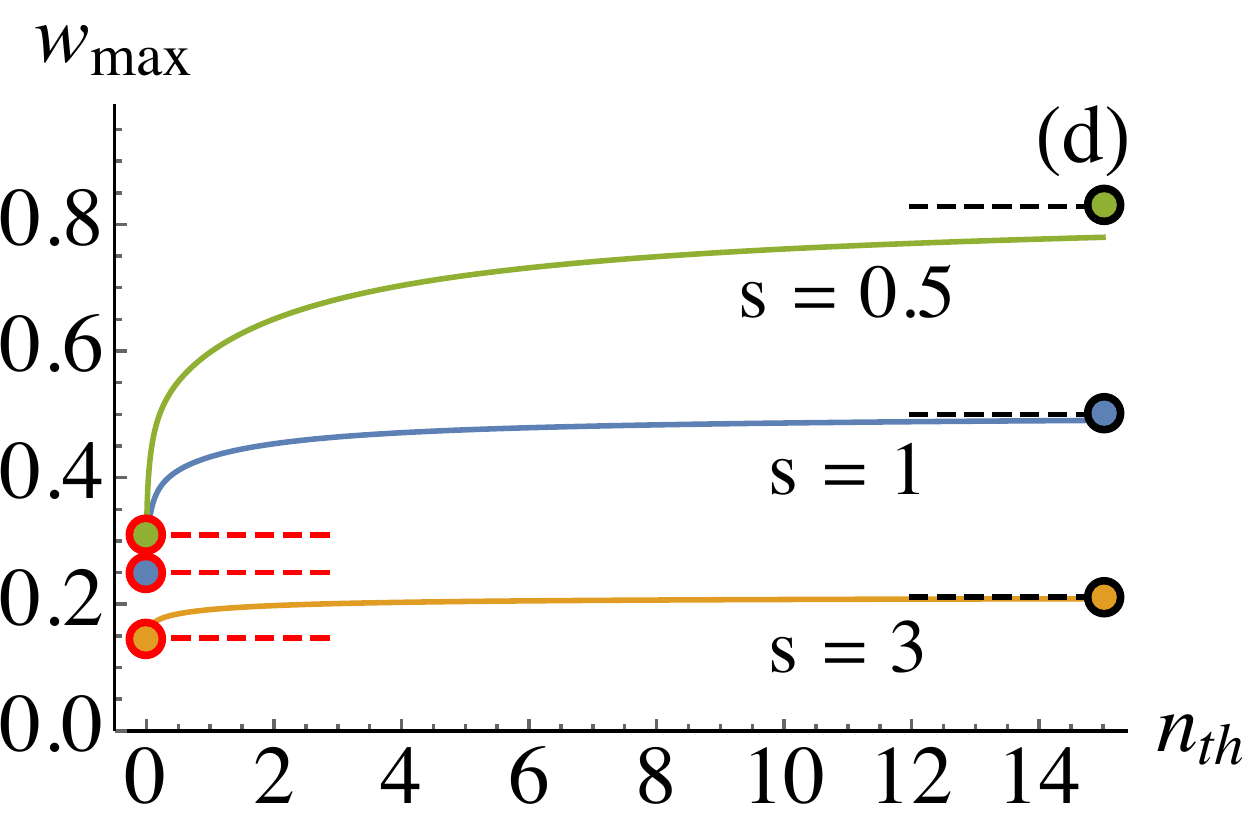}
\caption{(Color online) Plots of the sweet spots $\text{\it{w}}_\text{max}$ of the QFI as a function of (a) interaction time $ t$, (b) reservoir temperature $T$, (c) initial squeezing $\xi$ and (d) initial mean number of thermal excitations $n_\text{th}$. Each plot, obtained setting $\alpha=10^{-3}$, contains three curves corresponding to three types of structured environments ($s=1$, $s=3$ and $s=0.5$).  The highlighted red and black circles (together with constant dashed lines) indicate, respectively, the specific sweet spots $\text{\it{w}}^{(2)}_\text{max}$ and $\text{\it{w}}^{(4)}_\text{max}$ obtained by expanding the QFI for $\alpha\ll 1$ (see Sec.~\ref{s:ApprAlpha}). The fixed parameters are: (a) $k_B T=10^2 \omega_0$, $\xi=1$ and $n_\text{th}=2$; (b) $\omega_0 t=10^3$, $\xi=1$ and $n_\text{th}=2$; (c) $k_B T=10^2 \omega_0$, $n_\text{th}=2$ and $\omega_0 t=10^3$; (d) $k_B T=10^2 \omega_0$, $\xi=1$ and $\omega_0 t=10^3$. } \label{f:QFI_wmax_s}
\end{figure}
The main feature of these plots is that there exists always a maximum in each curve, corresponding to $\text{\it{w}}_\text{max}$, which we interpret as a ``sweet spot'' towards which the probe natural frequency $\omega_0$ has to be tuned in order to obtain the best estimation of the cutoff frequency $\omega_c$. Due to the importance of this result, it is useful to study the sweet spots as a function of the involved parameters (see Fig.~\ref{f:QFI_wmax_s}).
As it is clear from Fig.~\ref{f:QFI_wmax_s}, the sweet spot position $\text{\it{w}}_\text{max}$ depends on all the involved parameters and, in particular, it decreases for increasing values of interaction time $t$, temperature $T$ and initial squeezing $\xi$, whereas it increases for higher thermal excitation number $n_\text{th}$ in the initial state of the probe. By comparing all these plots, there are specific sweet spots (highlighted circles in Fig.~\ref{f:QFI_wmax_s}) that seem to depend only on the type of reservoir (parameter $s$). Indeed, as it will be discussed later on, by applying a series expansion of the QFI for $\alpha\ll 1$ under the condition of restricting the set of parameters, the positions of the sweet spots are, strikingly, only $s$-dependent and can be analytically predicted.
\par
\begin{figure}[t!]
\center
\includegraphics[width=0.237\textwidth]{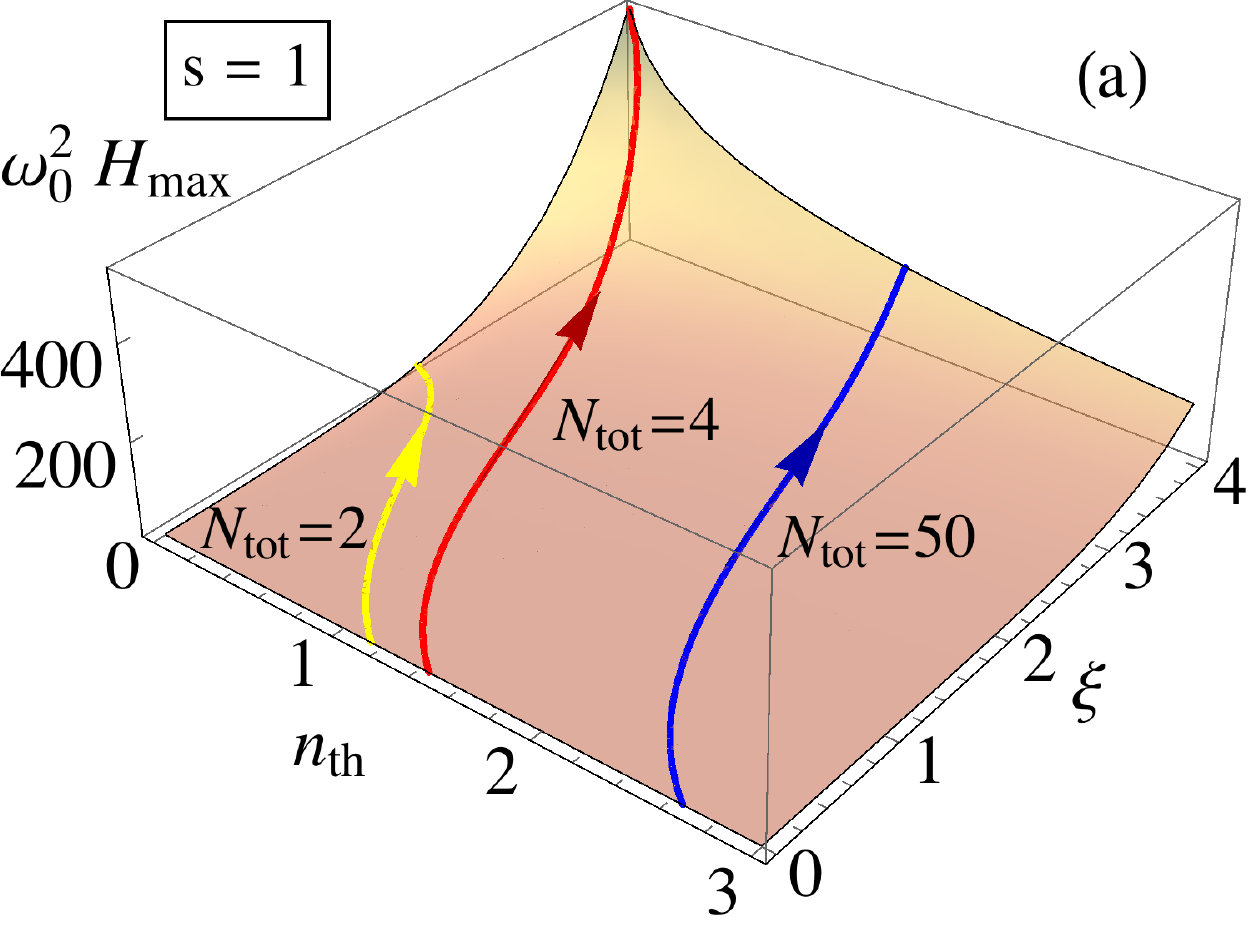}
\includegraphics[width=0.237\textwidth]{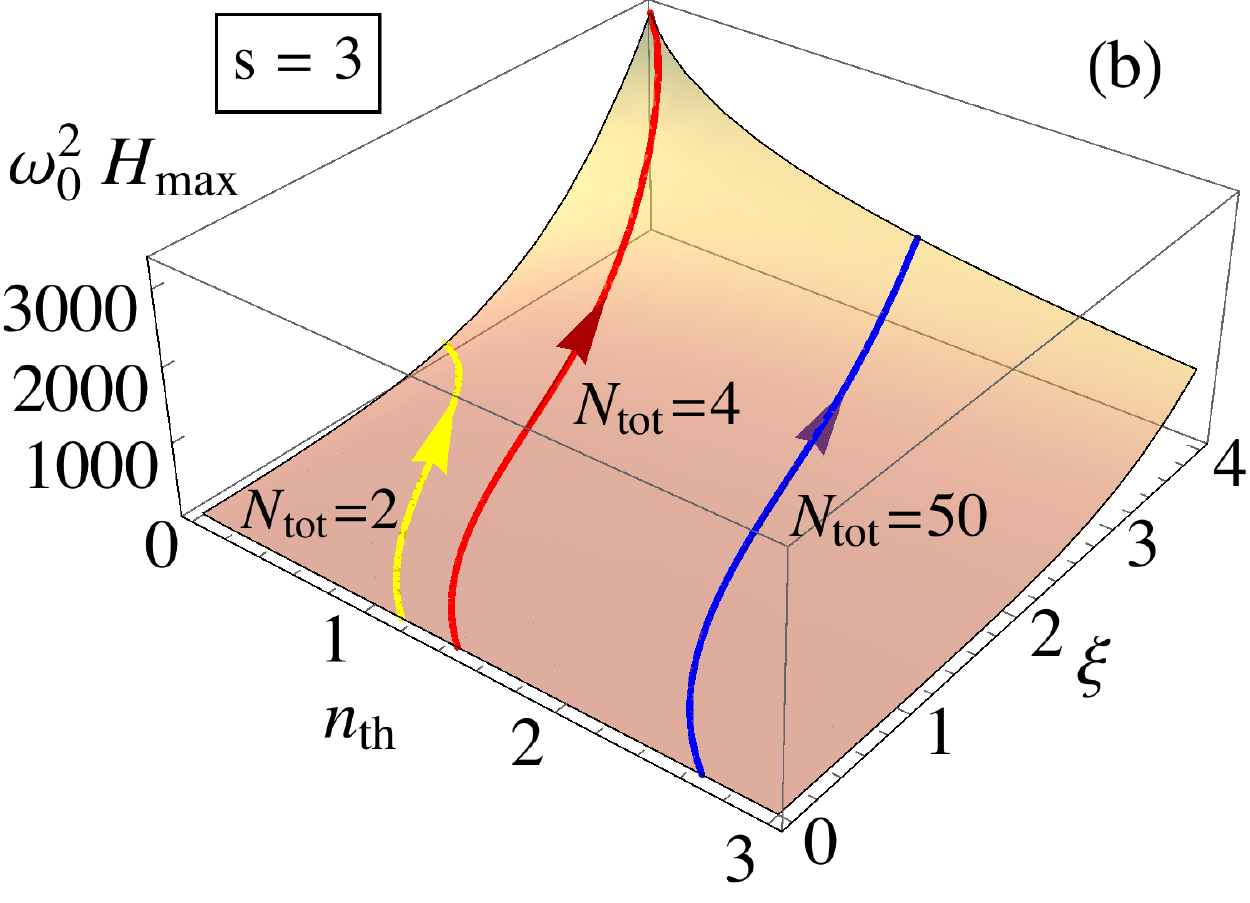}
\includegraphics[width=0.237\textwidth]{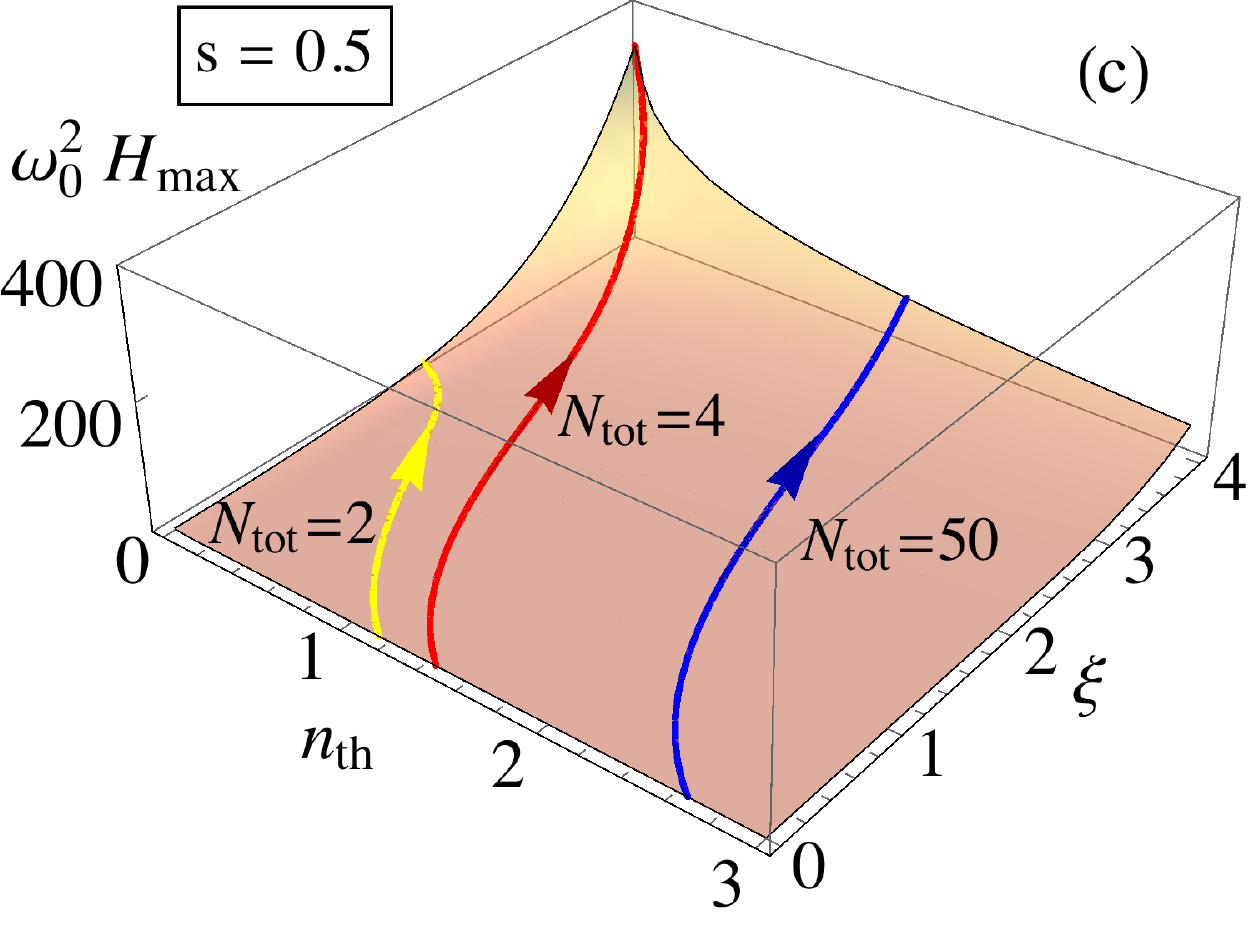}
\caption{(Color online) Plots of $H_\text{max}\equiv H(\text{\it{w}}_\text{max})$ as a function of the initial state parameters $\xi$ and $n_\text{th}$ for three cases of structured environments, (a) $s=1$, (b) $s=3$ and (c) $s=0.5$. The curves at fixed $N_\text{tot}$, mean total energy of the initial probe state, are highlighted by arrows in the direction of increasing $\xi$ and decreasing $n_\text{th}$, pointing towards the optimal squeezed vacuum state. The other parameters have been set to $\omega_0 t=10^3$, $k_B T= 10^2\omega_0$ and $\alpha=10^{-3}$.} \label{f:QFImax_nth_xi}
\end{figure}
In order to proceed with the optimization suggested by the estimation scheme in Fig.~\ref{f:schema}, we fix the environment parameters, and vary the initial conditions on the probe state preparation. Due to the presence of privileged frequencies (sweet spots) of the quantum harmonic oscillator used to probe the structured environment, the idea is to study the behavior of the maximum QFI, $H_\text{max}\equiv H(\text{\it{w}}_\text{max})$, as a function of initial squeezing $\xi$ and mean number of thermal excitations $n_\text{th}$, by tuning $\omega_0$ at the corresponding sweet spot $\text{\it{w}}_\text{max}$. In Fig.~\ref{f:QFImax_nth_xi}, we show the quantity $H_\text{max}$ for the three structured environment considered so far, namely $s=1$, $s=3$ and $s=1/2$. Apart from small differences and scaling values, the common trend in the plots of Fig.~\ref{f:QFImax_nth_xi} is that the optimal preparation of the probe state is the squeezed vacuum state and that the increase of squeezing enhances the estimation performances. This behavior can be better appreciated by fixing the mean total energy of a generic initial STS
\begin{equation}
N_\text{tot}\equiv \langle a^\dag a \rangle_{\varrho_0}= n_\text{th} + n_\text{sq}(2n_\text{th}+1) \, ,
\end{equation}
where the squeezing energy is $n_\text{sq}=\sinh^2 \xi$ and the thermal energy is $n_\text{th}$. The curves at constant $N_\text{tot}$ are highlighted by arrows in Fig.~\ref{f:QFImax_nth_xi}, where it is evident that $H_\text{max}$ increases whenever the squeezing contribution dominates over the thermal noise, up to the optimal squeezed vacuum state.
With this result, we demonstrate that the non-classicality of a squeezed state is a resource for the probing and the parameter estimation of structured environments.

\subsection{FI and performances of homodyne detection}\label{s:FI}
Having set the benchmarks for the optimization of the QFI, the next step is to find a feasible measurement scheme able to achieve the best possible estimation of the parameter of interest, i.e. the optimal FI.
We consider a homodyne detection scheme, where the corresponding observable is the generic quadrature $X(\varphi)$ on the probe evolved state $\varrho(t)$. The probability function $p(x| \lambda)$ of the parameter-dependent outcomes $x$ of the observable $X(\varphi)$ (see Eq.~(\ref{FI})), can be obtained as the marginal distribution in the real variable $x$ of a suitably transformed Wigner function:
\begin{equation}\label{Px_Wigner}
p(x|\lambda)=\int_{\mathbb{R}} \text{d}p\, \mathcal{W}[\varrho(t)](x \cos\varphi - p\sin\varphi, x\sin\varphi + p\cos\varphi)\, .
\end{equation}
Given the Gaussian nature of the probe state, the probability function $p(x|\lambda)$ is a normal distribution with zero mean and variance given by
\begin{equation}
\sigma_F = \sigma_{11} \cos^2\varphi+\sigma_{22}\sin^2\varphi+\sigma_{12} \sin2\varphi \, ,
\end{equation}
where $\sigma_{ij}$ are the elements of the CM (\ref{CM_Secular}) in the interaction picture, i.e. by considering $R(t)\rightarrow \mathbb{I}_{2\times 2}$. The FI (\ref{FI}) in the Markovian limit (\ref{Markov}), for a generic spectral density $J_\lambda(\omega)$, is given by
\begin{equation}\begin{split}
F(\lambda)=\frac{\dot{\sigma}_F^2}{2\sigma_F^2}= \left[\frac{c(\varphi) - \coth \big(\frac{\omega _0}{2k_B T}\big)} {c(\varphi)+2\,{\rm e}^{-\Gamma_M(t)} \Delta_{\Gamma_M}(t) }\right ]^2  \frac{\dot{\Gamma}^{\,2}_M(t)}{2}   \label{FI_X} \\ {}
\end{split}\end{equation}
where we defined 
\begin{equation}
c(\varphi) \equiv (1+2 n_\text{th})\left [\cosh 2\xi + \cos(2\varphi - \theta)\sinh 2\xi \right ] \, .
\end{equation}
\begin{figure}[t!]
\includegraphics[width=0.34\textwidth]{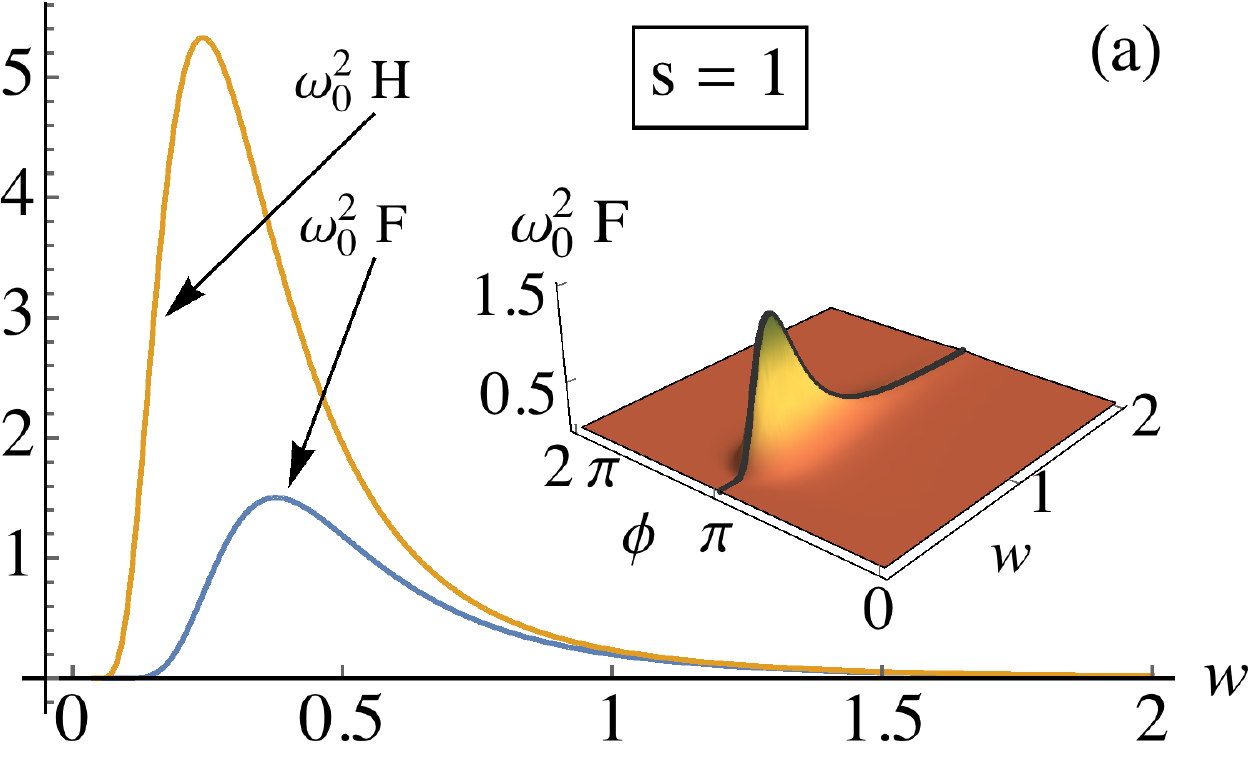}\\
\vspace{3mm}
\includegraphics[width=0.34\textwidth]{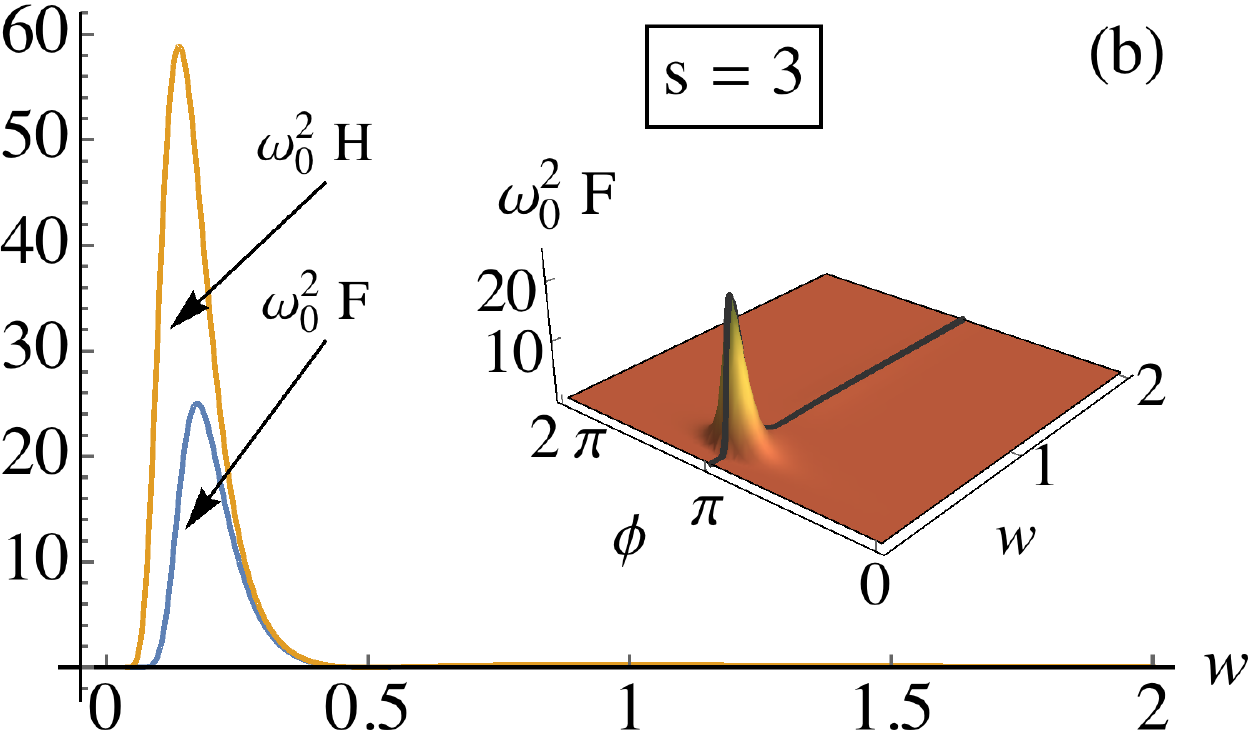}\\
\vspace{3mm}
\includegraphics[width=0.34\textwidth]{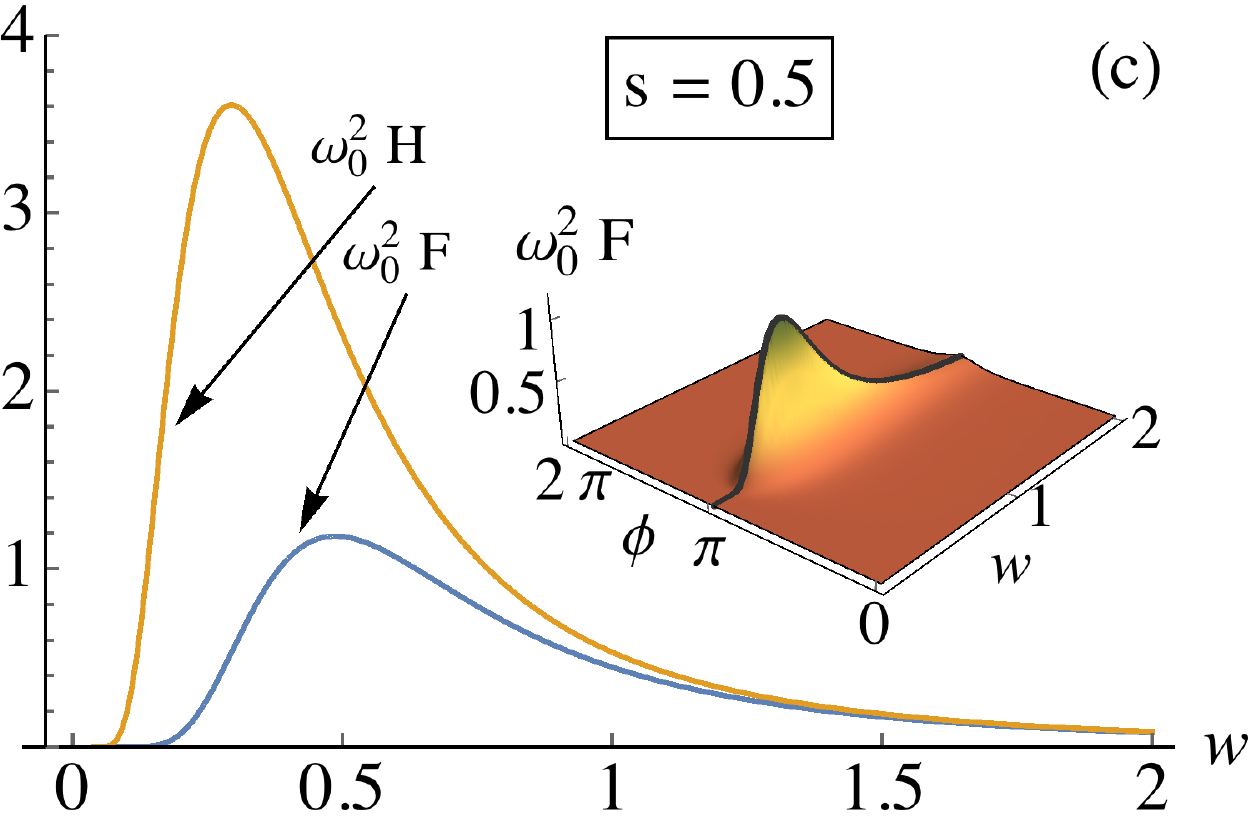}
\caption{(Color online) Plots of the FI of a homodyne measurement, with an initial squeezed vacuum state ($n_\text{th}=0$) of the probe, as a function of $\text{\it{w}}=\omega_c/\omega_0$ and a phase variable $\phi\equiv 2\varphi-\theta$ (see insets). The highlighted maximal curve at $\phi=\pi$ is also reported in the 2D-plots and compared to the corresponding QFI $\omega_0^2 H$ (solid curves). We considered three cases of structured environments, (a) $s=1$, (b) $s=3$ and (c) $s=0.5$, whereas the other parameters have been set to $\xi=1$, $\omega_0 t=10^3$, $k_B T=10^2 \omega_0$ and $\alpha=10^{-3}$.} \label{f:FI_wc_phi}
\end{figure}
Considering now the particular family of Ohmic spectral densities (\ref{J(w)}), we define $\phi\equiv 2\varphi-\theta$ and show in the insets of Fig.~\ref{f:FI_wc_phi} that the largest values of the FI are found in correspondence to $\phi=\pi$, i.e. by measuring the optimal quadrature $\varphi_\text{opt}=\frac{\theta+\pi}{2}$. This important result states that the best performances of a homodyne scheme are obtained by detecting the mode quadrature which has been mostly squeezed in the initial preparation of the probe state (minimum initial variance). We deduce that the non-classicality in the probe initial state provided by the squeezing, is a resource for the estimation strategy. Once the optimal quadrature to be measured is fixed, the FI as a function of the ratio $\text{\it{w}}=\omega_c/\omega_0$, displays a peak in correspondence of a sweet-spot $\text{\it{w}}_\text{max}$, located at a different position with respect to that of the QFI. This situation is evident for an initial squeezed vacuum state of the probe, for each of the three choices of structured environment (Ohmic, super-Ohmic and sub-Ohmic) we considered in Fig.~\ref{f:FI_wc_phi}. We also note that the FI saturates the QFI in the high-frequency range, thus reaching the QRCB at the expense of lowering the FI absolute values. As soon as the probe initial state is affected by thermal noise, i.e. by considering an initial STS, the situation is quite different and the two curves tend to be identical, as the thermal number of excitations increases. In the case of high thermal noise the homodyne scheme is quasi-optimal, in the sense that the ratio $r\equiv F/H \simeq 1$, but the cost to be paid is that the absolute values of FI (and QFI), get lower, which means a worse parameter estimation. 
\par
\begin{figure}[t!]
\includegraphics[width=.45\textwidth]{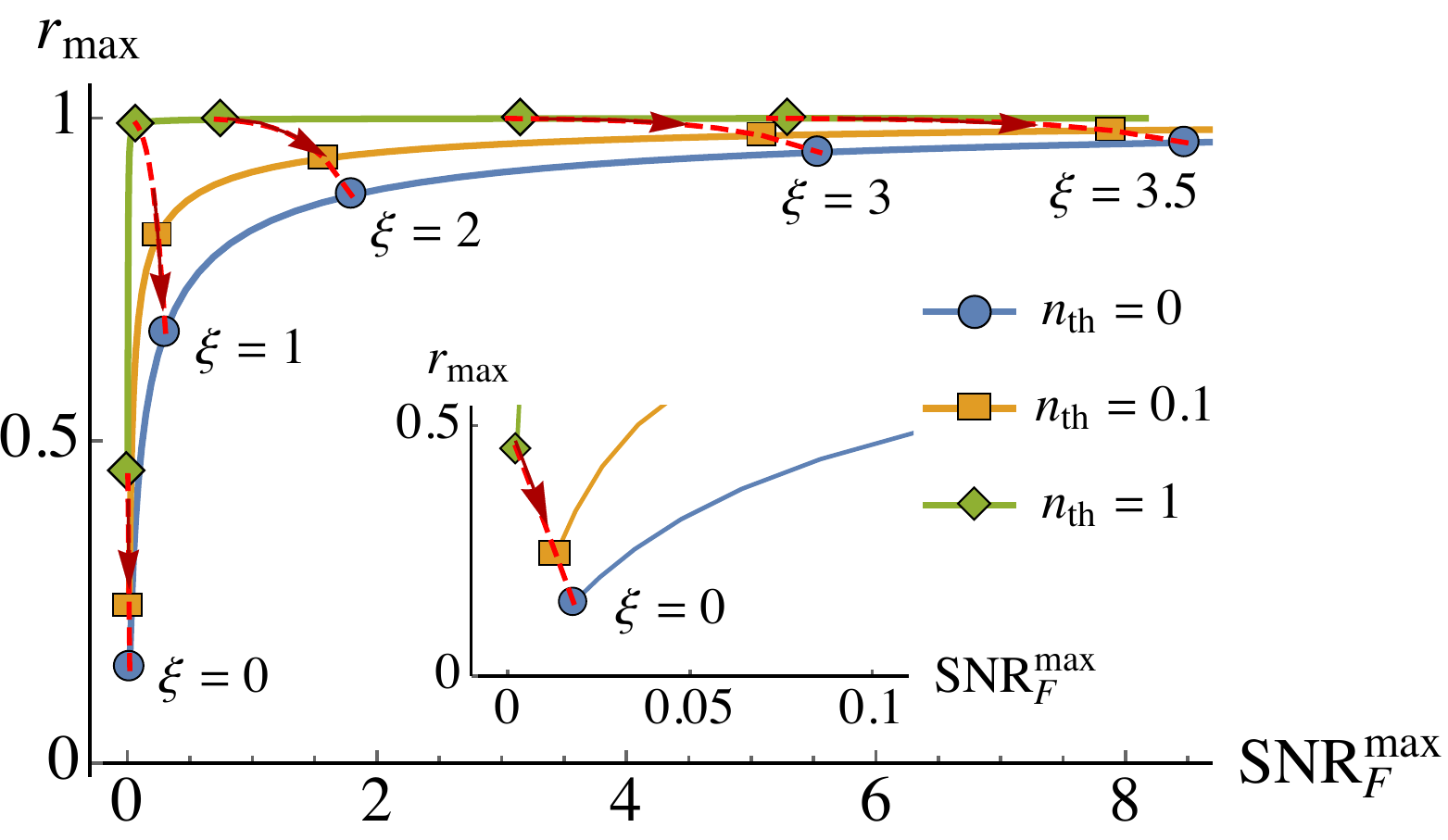}
\caption{(Color online) Trade-off between the maximum values of the SNR for the FI, $\text{SNR}_F^\text{max}$, and $r_\text{max}$, ratio between FI and QFI both evaluated at the sweet spot $\text{\it{w}}_\text{max}$ of $\text{SNR}_F$. Solid curves are obtained for three fixed values of the number of thermal excitations, $n_\text{th} =0, 0.1, 1$ and for squeezing parameter $\xi\in[0,4]$, whereas dashed red curves are obtained by fixing $\xi=0,1,2,3, 3.5$ and decreasing $n_\text{th}$. The solid blue curve with points represents the probe initialized in a squeezed vacuum state, while in the inset the probe is initialized as a pure thermal state, without squeezing. } \label{f:FIQFI0_FImax}
\end{figure}

There exists, then, a trade-off between our two main results: (i) maximization of the FI, or even better the $\text{SNR}_F$ (see Eq.~\ref{SNRFH}) by tuning the probe natural frequency at the specific sweet spot $\text{\it{w}}_\text{max}$ and (ii) saturation of the QRCB, i.e. the optimization of the FI with respect to the QFI. In Fig.~\ref{f:FIQFI0_FImax} we plot this trade-off between the maximum of the $\text{SNR}_F$, namely $\text{SNR}_F^\text{max}\equiv \omega_c^2F(\text{\it{w}}_\text{max})$, and the ratio $r_\text{max}\equiv F(\text{\it{w}}_\text{max})/H(\text{\it{w}}_\text{max})$. The curves are obtained for different values of the initial parameters $\xi$ and $n_\text{th}$, maximizing the $\text{SNR}_F$ with respect to the optimal quadrature $\varphi_\text{opt}$ and the corresponding sweet spot $\text{\it{w}}_\text{max}$, by fixing the other parameters such as time and reservoir temperature. The plot in Fig.~\ref{f:FIQFI0_FImax} can be read in two ways: firstly, at fixed number of thermal excitations, $\text{SNR}_F^\text{max}$ and $r_\text{max}$ increase for higher values of the initial squeezing parameter $\xi$ (solid curves), secondly, at fixed squeezing, $r_\text{max}$ increases for higher thermal component in the initial state of the probe, whereas $\text{SNR}_F^\text{max}$ decreases (dashed curves with arrows). This plot clearly shows that for an initial squeezed vacuum state the homodyne scheme is highly advantageous to obtain the best possible precision in estimating the cutoff frequency $\omega_c$ (high values of $\text{SNR}_F^\text{max}$), even though the FI does not reach the QCRB. Nonetheless, the squeezed vacuum state is difficult to obtain in an experiment and our results predict that, in the presence of thermal noise in the preparation of the probe, a homodyne measurement performs optimally as the FI reaches the QRCB, $r_\text{max}\simeq 1$, for moderately high squeezing and thermal excitation.

\section{More general Gaussian probes} \label{s:discussion}
In the previous Sections, we have analysed the conditions for the optimal estimation strategy 
of the cutoff frequency for probes prepared in a STS. In the following, we discuss the results 
for a generic initial Gaussian state of the probe. 
A generic single-mode Gaussian state can be written as $\varrho_0=D(\beta)S(\xi)\nu(n_{\text{th}})S^\dag(\xi)D^\dag (\beta)$, thus named displaced squeezed thermal state (DSTS). The coherent contribution of the state $\varrho_0$ is given by the displacement operator $D(\beta)=\exp\{\beta a^\dag-\beta^* a\}$, with $\beta=|\beta|{\rm e}^{\imm \theta_D}$, which coherently shifts a STS, centered at the origin of the phase space, of an amount $|\beta|$ in the direction $\theta_D$. In this case, both the initial and evolved first-moment vectors are no longer null, in particular $\vec{\delta}={\rm e}^{-\frac{\Gamma_M(t)}{2}} \vec{\delta}_0$. Considering a generic spectral density $J_\lambda(\omega)$, the employment of DSTSs brings an additional term to the expression of the QFI (\ref{QFIfinal}), given by
\begin{equation}\label{QFIadd}
\dot{\vec{\delta}}^{\,T} \sigma^{-1}\dot{\vec{\delta}}= n_c \frac{\Upsilon - (1+2n_\text{th})\cos\phi\sinh2\xi }{ 
    (1+2n_\text{th})^2+\Upsilon {\rm e}^{-\Gamma_M(t)} \Delta_{\Gamma_M}(t)}\dot{\Gamma}_M^{2}(t) \, ,
\end{equation}
where we defined the coherent energy $n_c\equiv |\beta|^2$, the angle $\phi\equiv 2\theta_D-\theta$ and the factor 
\begin{equation*}
\Upsilon \equiv (1+2n_\text{th})\cosh2\xi +2\,{\rm e}^{-\Gamma_M(t)} \Delta_{\Gamma_M}(t) \, .
\end{equation*}
We remember that we are considering a Markov regime and that the derivative in Eq.~(\ref{QFIadd}) is meant with respect to the parameter of interest $\lambda\to\omega_c$.
\begin{figure}[t!]
\includegraphics[width=.237\textwidth]{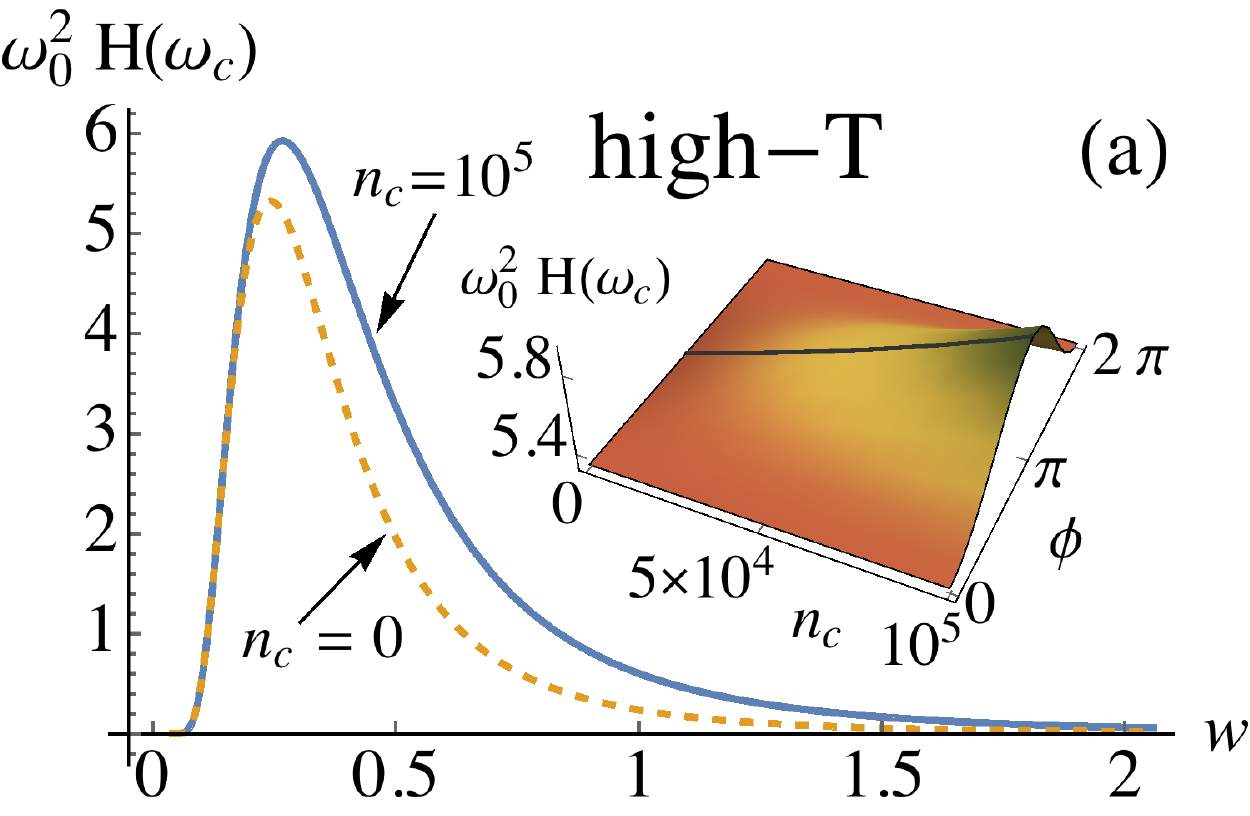}
\includegraphics[width=.237\textwidth]{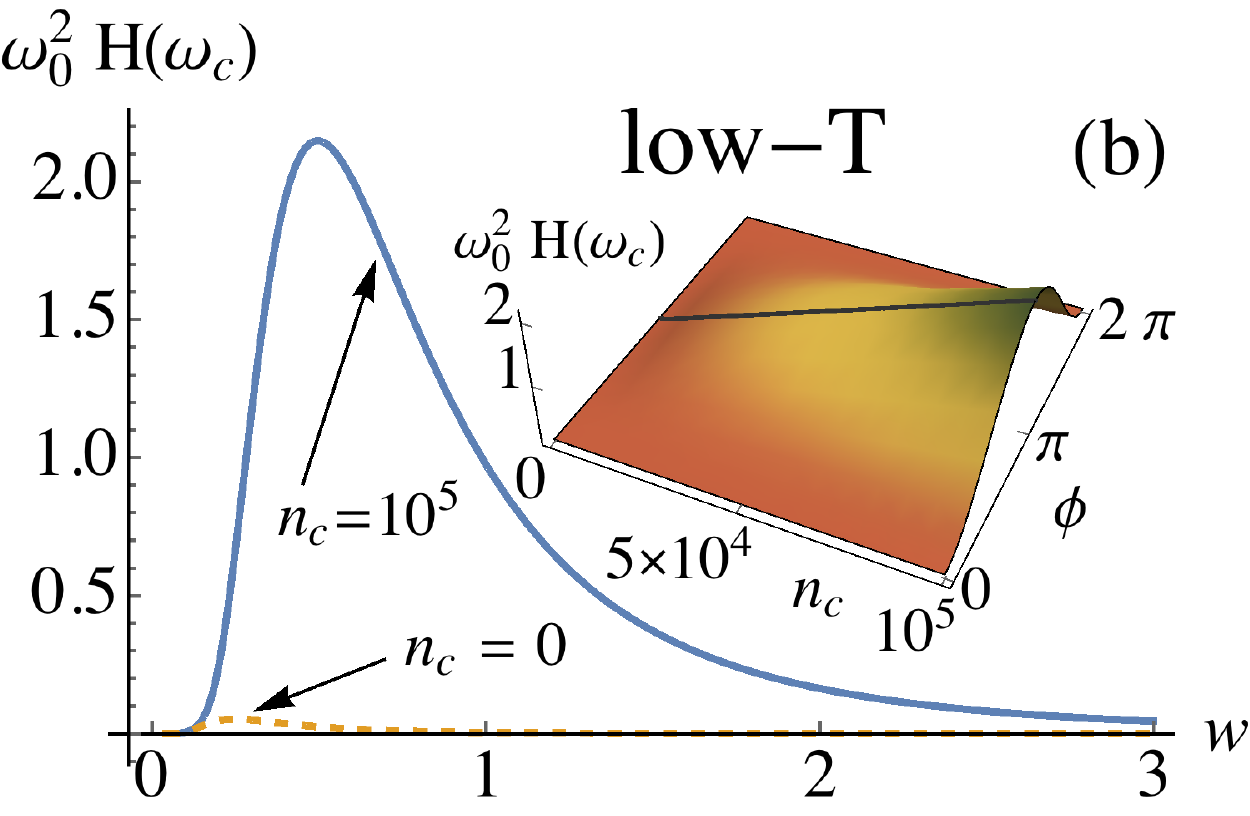}
\caption{(Color online) Plots of $H(\omega_c)$, rescaled by $\omega_0^2$, as a function of $\text{\it{w}}=\omega_c/\omega_0$ for a probe initially prepared in a displaced squeezed vacuum state, with $n_c=10^5$ (solid curve), compared with the case of an initial squeezed vacuum state, with $n_c=0$  (dashed curve). The optimal direction of displacement is $\theta_{D,\text{opt}}=\frac{\theta+\pi}{2}$, which maximizes the QFI, as show in the inset. The values of the employed parameters are $\xi=1$, $\omega_0 t=10^3$ and $\alpha=10^{-3}$, whereas in (a) $k_B T=10^2 \omega_0$ and in (b) $k_B T= \omega_0$. The choice of the Ohmic spectral density $s=1$ is exemplifying of the other types of environments. } \label{f:DSTS_H}
\end{figure}
\par
\begin{figure}[t!]
\center
\includegraphics[width=0.237\textwidth]{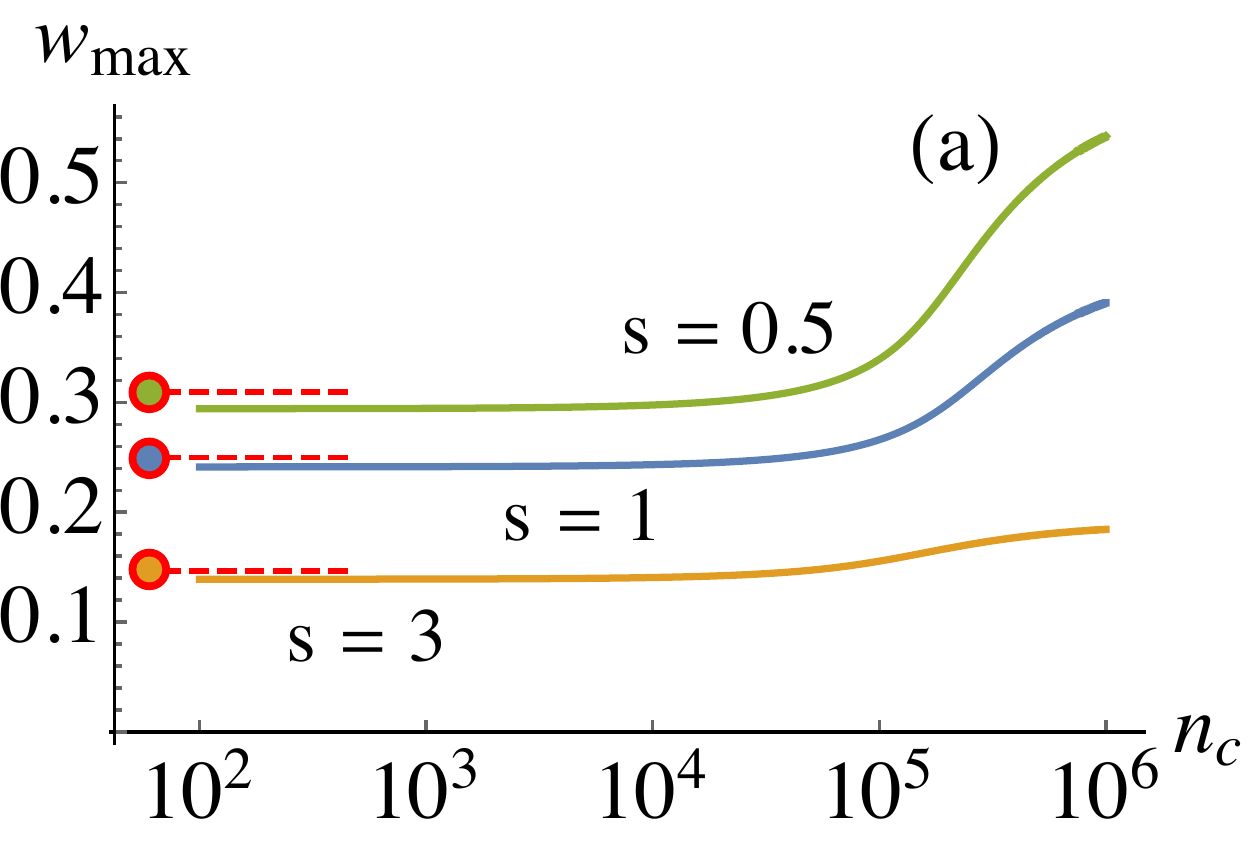}
\includegraphics[width=0.237\textwidth]{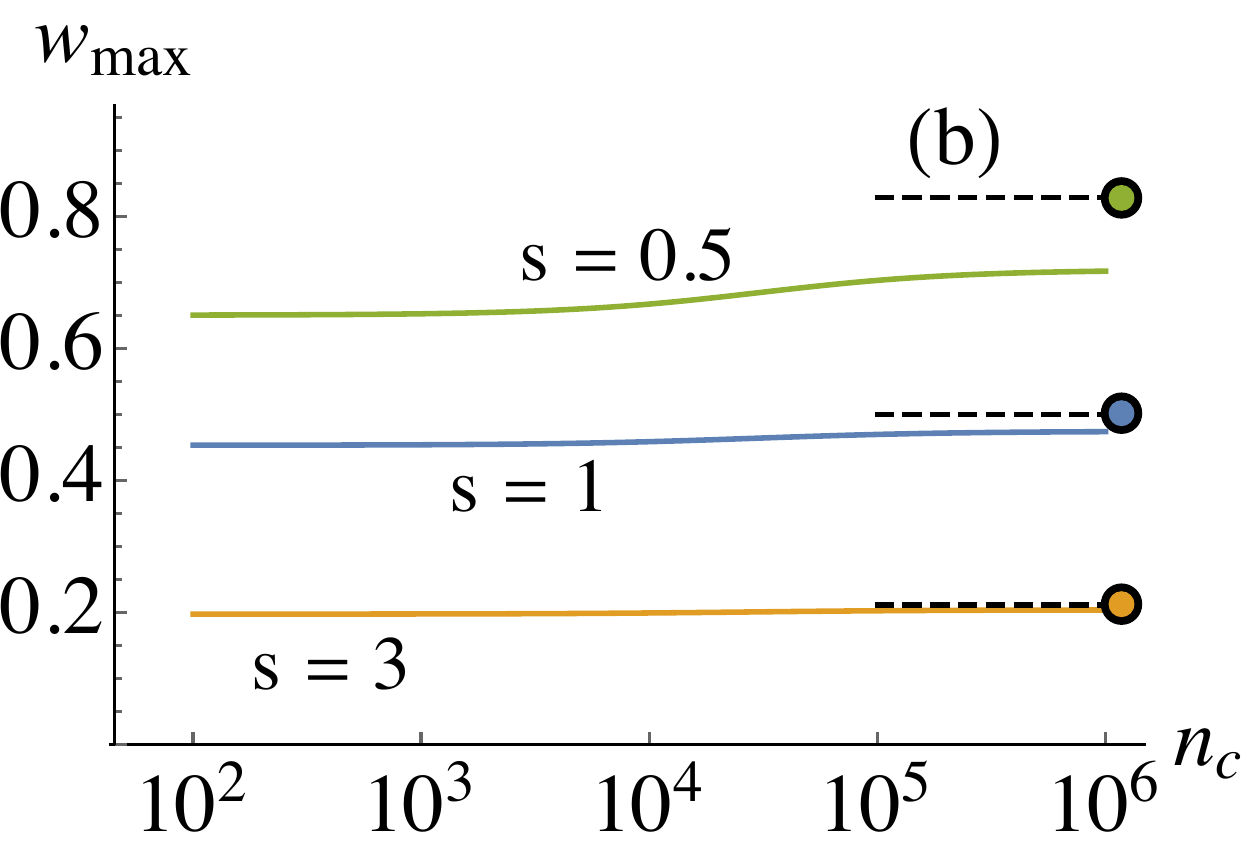}\\
\includegraphics[width=0.237\textwidth]{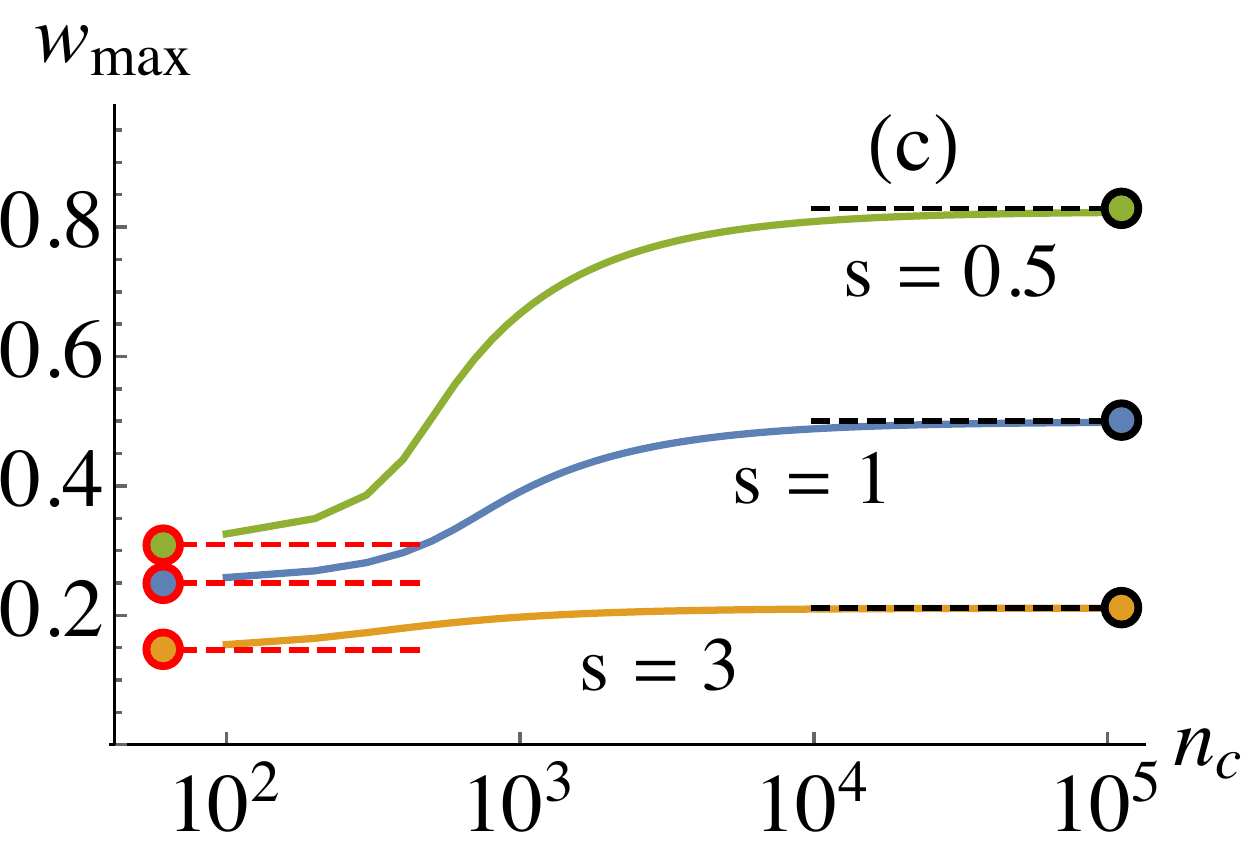}
\includegraphics[width=0.237\textwidth]{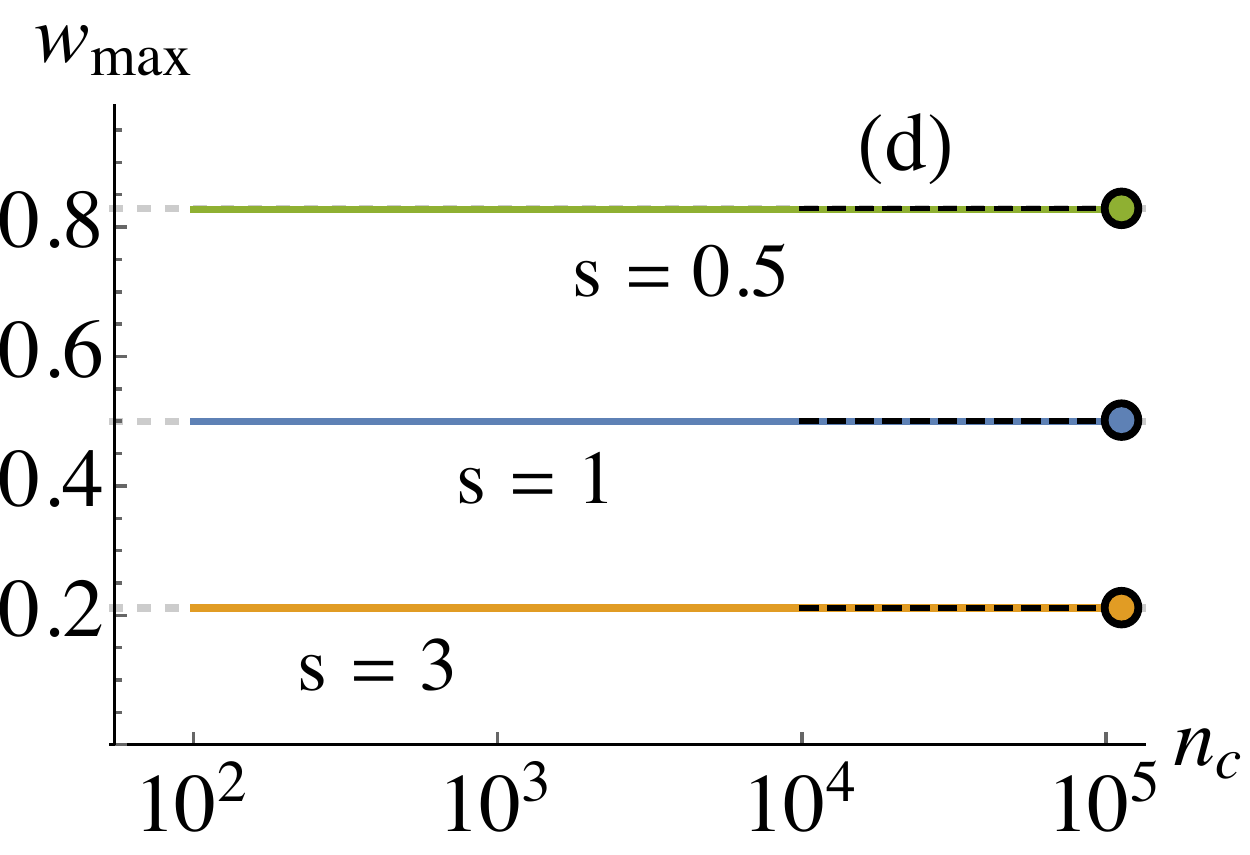}
\caption{(Color online)  Plots of the sweet spots $\text{\it{w}}_\text{max}$ of the QFI as a function of the coherent input energy $n_c$, maximized at $\theta_{D,\text{opt}}$ as described above, for an initial DSTS of the probe with $n_\text{th}=0$ (a)-(c) and $n_\text{th}=2$ (b)-(d) and for two regimes of temperature, $k_B T=10^2 \omega_0$ (a)-(b) and $k_B T=10^{-2} \omega_0$ (c)-(d). The highlighted red and black circles (together with constant dashed lines) indicate the specific sweet spots $\text{\it{w}}^{(2)}_\text{max}$ and $\text{\it{w}}^{(4)}_\text{max}$, respectively, under a series expansion for $\alpha\ll 1$ applied to the QFI. Note that for high temperatures (a-b) these specific sweet spots are never reached, as suggested by Fig.~\ref{f:QFI_wmax_s}(b). Each plot contains three curves corresponding to three types of structured environments ($s=1$, $s=3$ and $s=0.5$). The fixed parameters are: $\xi=1$, $\omega_0 t=10^3$,  and $\alpha=10^{-3}$.} \label{f:QFI_wmax_nc}
\end{figure}

The contribution (\ref{QFIadd}) to $H(\omega_c)$ depends linearly on the coherent energy $n_c$ and it is maximum for $\phi=\pi$ (see inset in Fig.~\ref{f:DSTS_H}), for the particular choice of the Ohmic spectral density. This intuitive and rather important result means that the angle of displacement maximizing the QFI, is the same of the initially mostly squeezed quadrature, i.e. $\theta_{D,\text{opt}}=\frac{\theta+\pi}{2}$. In Fig.~\ref{f:DSTS_H}(a) we show that the effect of displacement, for an initial state of the probe with zero thermal noise, is to increase the QFI and shift the correspondent sweet spot $\text{\it{w}}_\text{max}$ (the effect is more pronounced for low-temperature environments, Fig.~\ref{f:DSTS_H}(b)). The shifting of $\text{\it{w}}_\text{max}$ depends on the environment temperature $T$ and it is not pronounced for initial DSTS with non-zero thermal noise (Figs.~\ref{f:QFI_wmax_nc}(b)-(d)). When the initial state is a displaced squeezed vacuum (Figs.~\ref{f:QFI_wmax_nc}(a)-(c)), this effect is amplified and the coherent term allows to shift the sweet spots to the highest value allowed by a further weak-coupling approximation $\alpha\ll 1$ for the QFI (see Section \ref{s:ApprAlpha}). We point out that the plots in Fig.~\ref{f:QFI_wmax_nc} have an exaggerated axis scale of the coherent energy in order to display the effect of shifting the sweet spots, but for ordinarily small ranges of $n_c$ the positions of the sweet spots are basically fixed.
\par
\begin{figure}[t!]
\center
\includegraphics[width=0.237\textwidth]{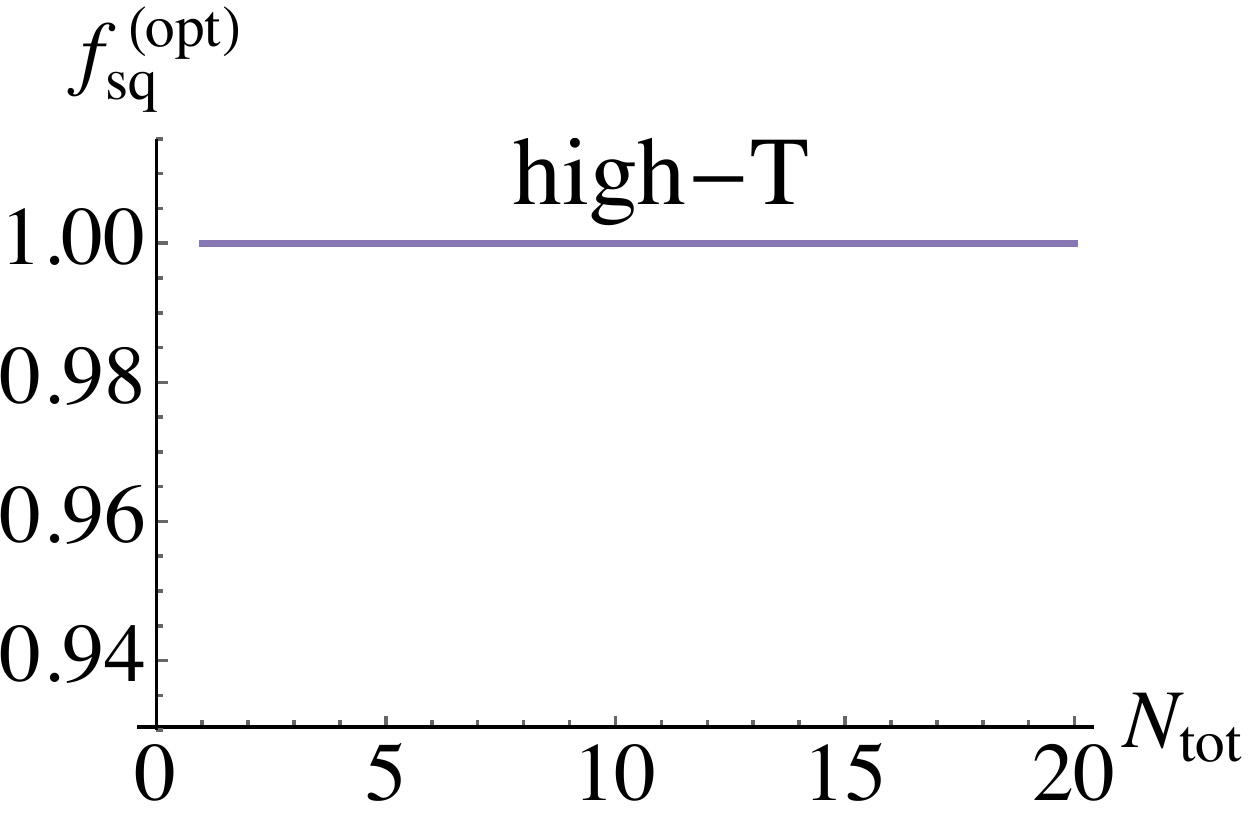}
\includegraphics[width=0.237\textwidth]{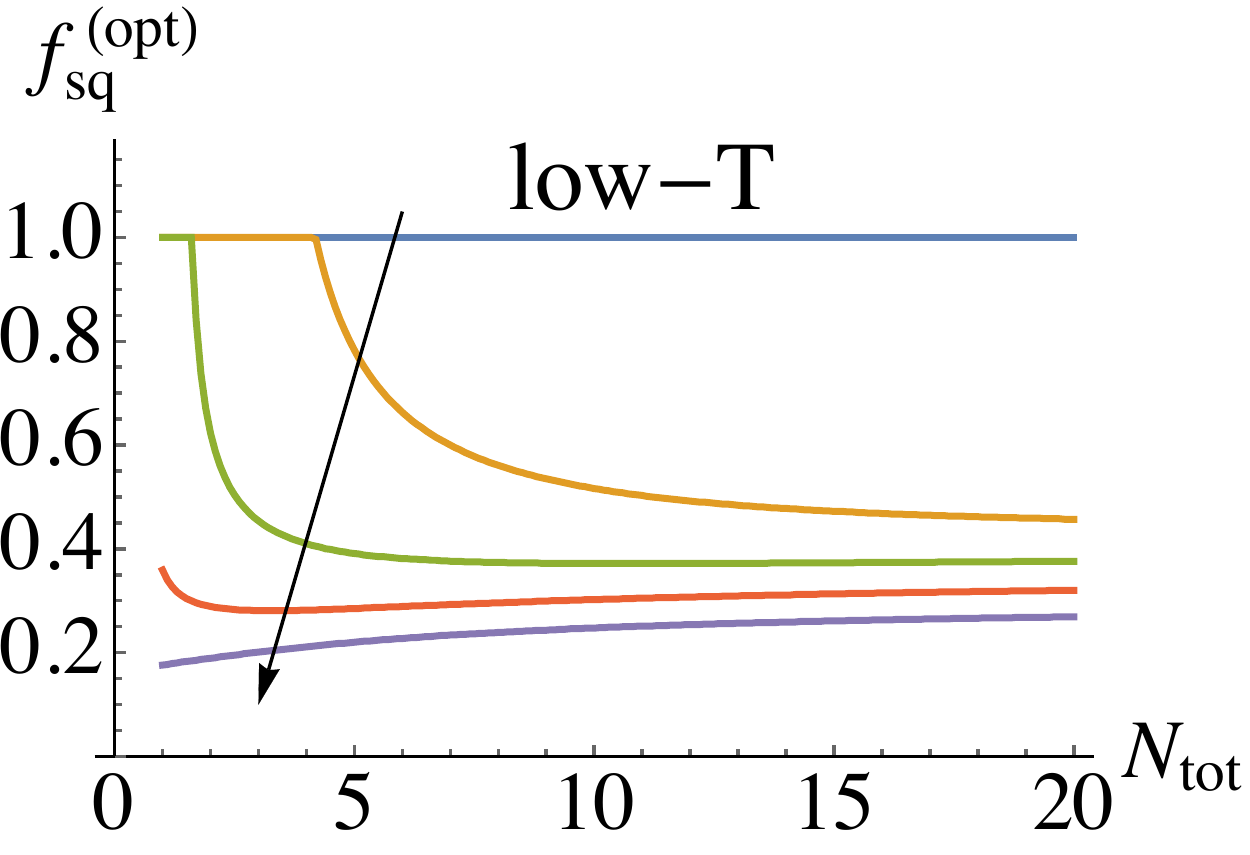}
\caption{(Color online) Plots of the optimal squeezing fraction $f_\text{sq}^{\text{(opt)}}$ as a function of the total energy $N_\text{tot}$ of the input probe DSTS, for increasing thermal fraction $f_\text{th}=0, 0.1, 0.2, 0.3, 0.4$ (as indicated by the arrow). For a high-temperature environment, $k_B T=10^2 \omega_0$, $f_\text{sq}^{\text{(opt)}}=1$ for all the considered ranges of total energy and thermal fractions. For a low-temperature environment, $k_B T=10^{-2} \omega_0$, $f_\text{sq}^{\text{(opt)}}\neq 1$ as long as $f_\text{th}\neq 0$ and $N_\text{tot}$ increases.} \label{f:fsOpt}
\end{figure}

In order to compare different initial states of the probe and to analyze the effects of displacement on the QFI, it is more convenient to re-parametrize the QFI in terms of the different energy contributions to the total energy of the initial state, $N_\text{tot}= n_\text{th} + n_\text{sq}(2n_\text{th}+1)+n_c$. Thus we define the squeezing fraction $f_\text{sq}$, the coherent fraction $f_c$ and the thermal fraction $f_\text{th}$, as follows:
\begin{subequations}\begin{align}
f_\text{sq}&\equiv \frac{\sinh^2\xi(1+2n_\text{th})}{N_\text{tot}-n_\text{th}}\\
f_c&\equiv 1-f_\text{sq} = \frac{n_c}{N_\text{tot}-n_\text{th}}\\
f_\text{th}&\equiv \frac{n_\text{th}}{N_\text{tot}} \, .
\end{align}\end{subequations}
We address the question of how much squeezing fraction $f_\text{sq}^{\text{(opt)}}$ or coherent fraction $f_c^{\text{(opt)}}$ should be employed in the initial state to maximize the QFI at a given thermal noise fraction and total amount of energy. In the absence of thermal noise $f_\text{th}=0$, the most convenient strategy is to employ all the energy of the initial state of the probe into squeezing, thus always obtaining $f_\text{sq}^{\text{(opt)}}=1$ for increasing total energy $N_\text{tot}$ and for all considered environment temperatures. Whenever the thermal fraction is kept fixed $f_\text{th}\neq 0$, the optimal value for the squeezing fraction is $f_\text{sq}^{\text{(opt)}}< 1$ for low environment temperatures, as $N_\text{tot}$ increases (see Fig.~\ref{f:fsOpt}). This worth result is really important from an experimental point of view. In fact, for increasing total energy and thermal fraction, the most convenient strategy is to employ the energy into displacement instead of squeezing, this last being commonly difficult to increase \cite{GenoniOlivares}.

\section{Weak coupling expansion of the QFI}\label{s:ApprAlpha}
Let us now focus on a further approximation, valid for $\alpha\ll 1$, applied to the QFI. We remark that the weak coupling is a reasonable approximation for a probing scheme, as a probe should not produce much disturbance on the environment. We thus consider the first non vanishing term in the expansion of $H(\lambda)$ for $\alpha\ll 1$, which contains functions of the coefficient 
$\Gamma(\lambda)\propto \alpha^2$ and it allows to obtain interesting analytic results. 
The first one is that if we consider an initial squeezed vacuum state ($n_\text{th}=0$), 
the expanded QFI is $H(\lambda) \propto \alpha^2$, while for $n_\text{th}\neq 0$ the first 
non-null term of the QFI is of the order $\alpha^4$. This approximated result, confirmed by numerical analysis of the complete QFI (\ref{QFIfinal}), naturally yields to the first consideration: a single-mode quantum harmonic oscillator probe, initialized in a squeezed vacuum state $\varrho_0=S(\xi)\ketbra{0}{0}S^\dag(\xi)$, is the optimal probe state for the quantum estimation of $\lambda$
\cite{gaiba}. The simplified expressions of the QFI at the order $\alpha^2$ for an initial squeezed vacuum state and at the order $\alpha^4$ for an initial STS are:
\begin{align}\label{QFiweak2}
&H^{(2)}(\lambda)=\left [\cosh(2\xi)\coth\left(\frac{\omega_0}{2k_B T}\right) -1 \right ] \frac{\dot{\Gamma}_M^2(t)}{2\, \Gamma_M(t)}\\
\begin{split}\label{QFiweak4}
&H^{(4)}(\lambda)=\Bigg[\frac{ \coth ^2\left(\frac{\omega_0}{2k_B T}\right) \left((1+2 n_\text{th})^2 \cosh (4 \xi )+1\right)}{(1+2n_\text{th})^4-1}\\
&-\frac{4 \big(2 n_\text{th} (1+n_\text{th})+1\big) (1+2n_\text{th}) \cosh (2 \xi ) \coth \left(\frac{\omega_0}{2k_B T}\right)}{(1+2n_\text{th})^4-1}\\
&+\frac{\big(4 n_\text{th} (1+n_\text{th})+2\big) (1+2n_\text{th})^2}{(1+2n_\text{th})^4-1}\Bigg ] \dot{\Gamma}_M^2(t) \, .
\end{split}
\end{align}
It is important to point out that, given the explicit dependence on the model parameters of Eqs.~(\ref{QFiweak2}, \ref{QFiweak4}), the approximation $\alpha\ll 1$ is valid if the other parameters are properly taken into account. In more detail, the behavior of the QFI is monotonically increasing with $\xi$, $T$ and $t$, while the QFI decreases with $n_\text{th}$. Thus, in order Eqs.~(\ref{QFiweak2}, \ref{QFiweak4}) to be valid, limited allowed ranges of parameters should meet together. For this reason, we preferred to present our main results in the previous Sections without performing perturbative expansions of the QFI. Nonetheless, Eqs.~(\ref{QFiweak2}, \ref{QFiweak4}) provide important insights on the estimability of the cutoff frequency.
\par
\begin{figure}[t!]
\center
\includegraphics[width=0.235\textwidth]{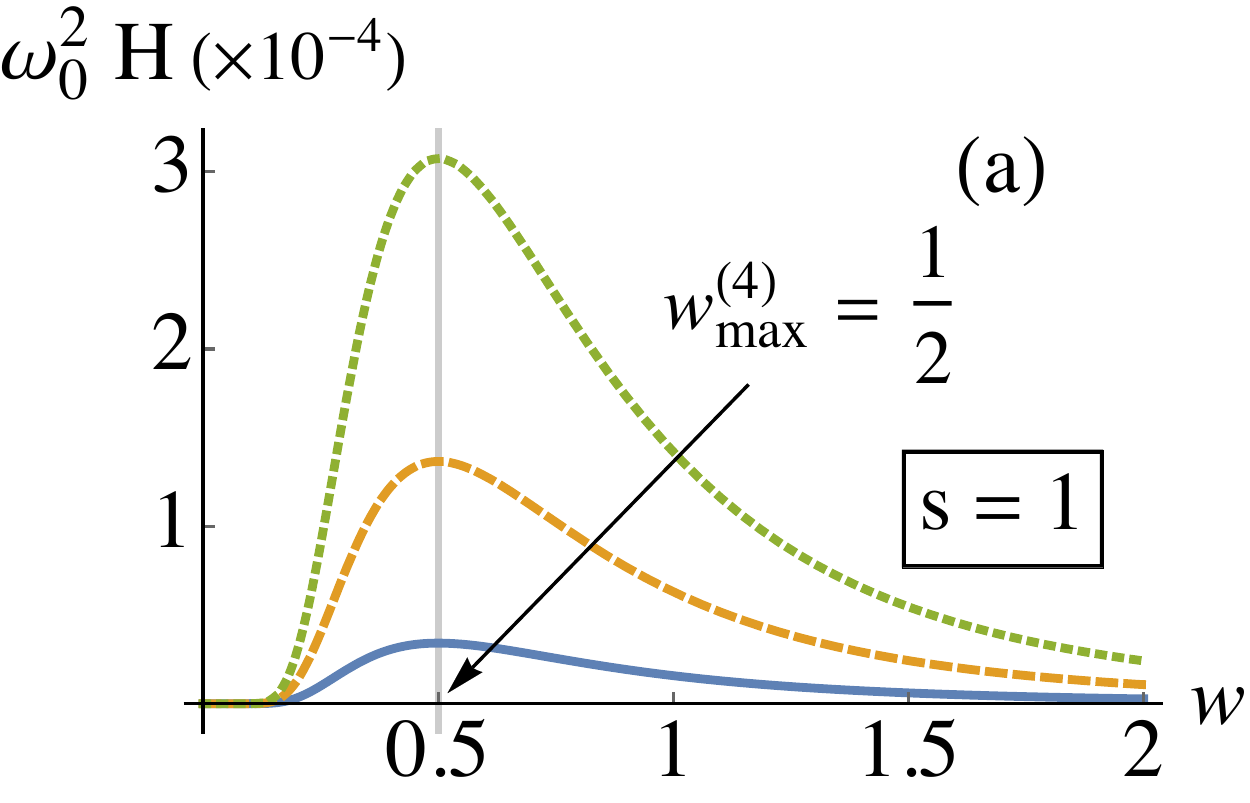}\hspace{1mm}
\includegraphics[width=0.235\textwidth]{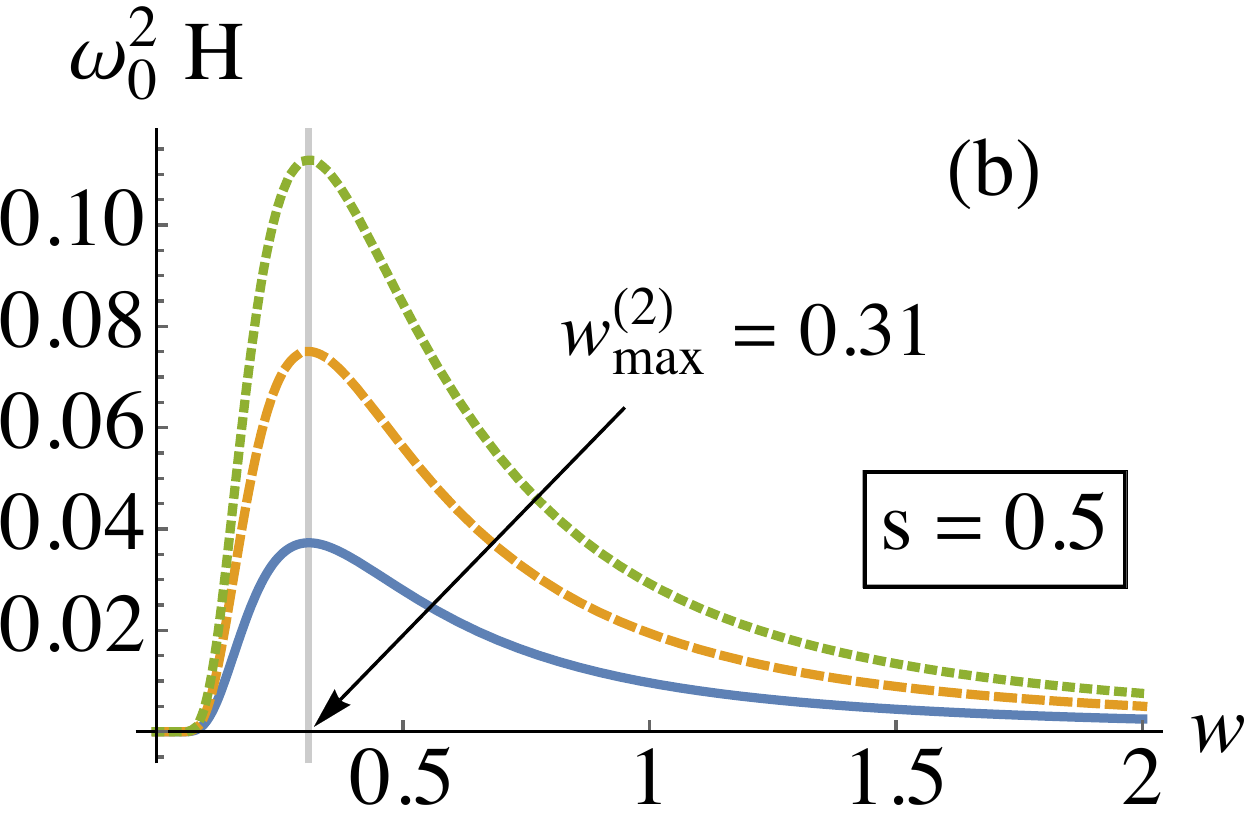}\\
\includegraphics[width=0.235\textwidth]{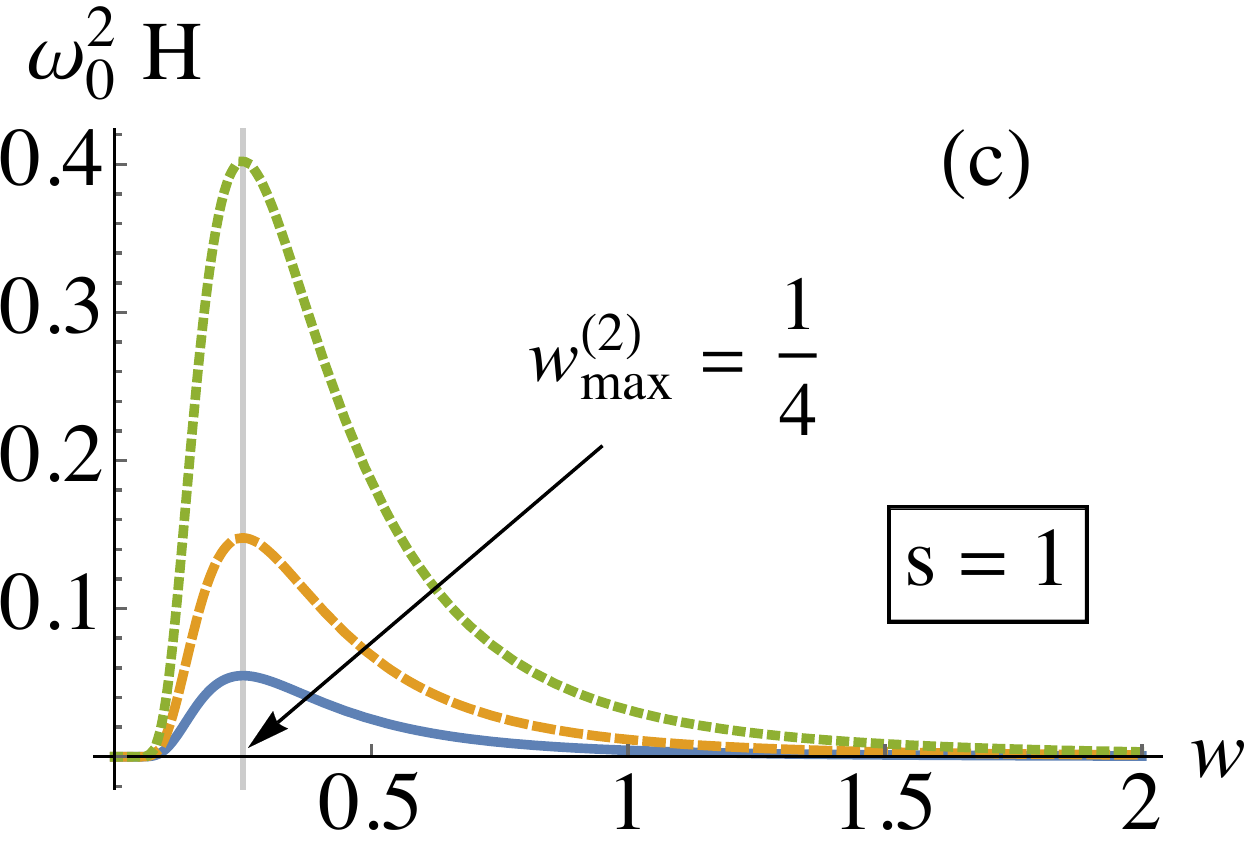}\hspace{1mm}
\includegraphics[width=0.235\textwidth]{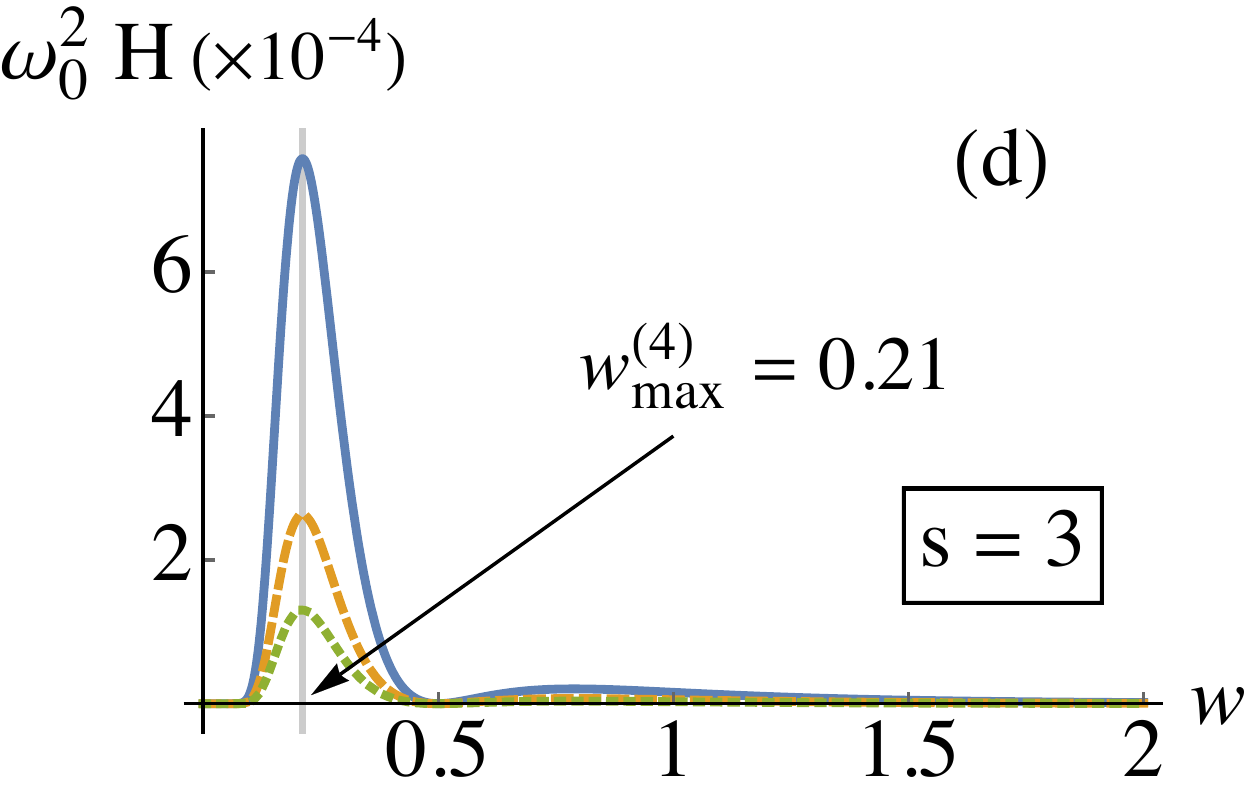}
\caption{(Color online) Plots of the QFI, rescaled by $\omega_0^2$, as a function of $\text{\it{w}}=\omega_c/\omega_0$ under the approximation $\alpha\ll 1$, specifically with $\alpha=10^{-3}$. The fixed sweet spot positions are given by Eqs.~(\ref{wmax2},\ref{wmax4}). (a) Increasing time: $\omega_0 t = 100$ (blue solid), $\omega_0 t = 200$ (orange dashed) and $\omega_0 t = 300$ (green dotted). Fixed parameters $s=1$, $k_B T=10\,\omega_0$, $\xi=1$ and $n_\text{th}=1$. (b) Increasing reservoir temperature: $k_B T=10\,\omega_0$ (blue solid), $k_B T=20\,\omega_0$ (orange dashed) and $k_B T=30\,\omega_0$ (green dotted). Fixed parameters $s=1/2$, $\omega_0 t=100$, $\xi=1$ and $n_\text{th}=0$.  (c) Increasing squeezing: $\xi =1$ (blue solid), $\xi = 1.5$ (orange dashed) and $\xi=2$ (green dotted). Fixed parameters $s=1$, $k_B T=10\,\omega_0$, $\omega_0 t=100$ and $n_\text{th}=0$. (d) Increasing initial thermal energy: $n_\text{th}=1$ (blue solid), $n_\text{th}=2$ (orange dashed) and $n_\text{th}=3$ (green dotted). Fixed parameters $s=3$, $k_B T=10\,\omega_0$, $\omega_0 t=100$ and $\xi=1$. } \label{f:QFIapprox_wmax_s}
\end{figure}

The most striking consequence of this further approximation $\alpha \ll 1$ is that the sweet spots, i.e. $\text{\it{w}}_\text{max}\equiv \omega_c/\omega_0$ maximizing the QFI, depend only on the type of structured environment. In the following the expression of $\text{\it{w}}_\text{max}$, only $s$-dependent, for the first non-null terms in the expansions (\ref{QFiweak2}) and (\ref{QFiweak4}):
\begin{align}\label{wmax2}
&\text{\it{w}}^{(2)}_\text{max}= \Bigg\{
               \begin{array}{rl}
                & \frac{s+1-\sqrt{2(s+1)}}{s^2 -1} \quad\mbox{for }\, s\neq 1  \\
                 & 1/4 \hspace{1.15cm}\mbox{for }\, s=1 \, ,
               \end{array} \\      
&\text{\it{w}}^{(4)}_\text{max}= \Bigg\{
               \begin{array}{rl}
                 &\frac{1}{s +\sqrt{s}} \quad\mbox{for }\, s\neq 1  \\
                  &1/2 \hspace{1.15cm}\mbox{for }\, s=1 \, .
               \end{array} \label{wmax4}       
\end{align}
Fixing the environment Ohmic parameter $s$, the value of $H(\omega_c)$ increases or decreases depending on which parameter is varied, as shown with some examples in Fig.~\ref{f:QFIapprox_wmax_s}. Ultimately, the QFI expansion for $\alpha\ll 1$ (Eqs.~(\ref{QFiweak2},\ref{QFiweak4})) identifies two limiting cases for the sweet-spot positions (Eqs.~(\ref{wmax2},\ref{wmax4})) determined by the initial state of the probe, squeezed vacuum or squeezed thermal states (highlighted, respectively, by red and black circles both in Fig.~\ref{f:QFI_wmax_s} and Fig.~\ref{f:QFI_wmax_nc}). We note that the sweet spot position of Eq.~(\ref{wmax4})  is valid also for the approximated expression of the FI in Eq. (\ref{FI_X}), independently of the initial state of the probe, as the first non-null term in the expansion of the FI is always of the order $\alpha^4$.
\par
This further weak-coupling expansion for $\alpha\ll 1$ provides, also, an analytic insight of the important result described in Sec.~\ref{s:FI}, i.e. the optimal quadrature to be measured in a homodyne scheme is the one which has been mostly squeezed in the initial state.
Let us consider the projectors on the eigenspaces of the SLD operator $\mathcal{L}_{\lambda}$, for a generic parameter of interest $\lambda$ of the spectral density $J_\lambda(\omega)$. They constitute the optimal POVM for the estimation of the parameter of interest $\lambda$, so that the FI associated to this measurement equals the QFI. For simplicity, we consider an initial STS for the probe and Eq. (\ref{SLDgaussian}) reduces to $\mathcal{L}_\lambda=\vec{Z}^{\,T}\Phi\vec{Z}$, with the real matrix
\begin{equation}  \label{Phi}
 \Phi=\frac{1}{2{d}^4-1/8}\left\lbrace
     {d}^4\,\sigma^{-1}\dot{\sigma}\sigma^{-1}
      -\frac{1}{4}\,\Omega\dot{\sigma}\Omega
      \right\rbrace  \, ,                                            
\end{equation}
where $\sigma$ is the covariance matrix of the Gaussian state, $d$ its symplectic eigenvalue and $\Omega$ is given by Eq.~(\ref{Omega}).
In order to compute (\ref{Phi}), we switch to the interaction picture, resulting in the substitution $R(t)\rightarrow \mathbb{I}_{2\times 2}$, with $R(t)$ given in Eq. (\ref{R(t)}). Performing the Markov approximation and the series expansion for $\alpha\ll 1$, we can easily diagonalize $\Phi = \mathsf{D}\,\Phi_{\text{diag}} \mathsf{D}^T $, with
\begin{align}
 \Phi_{\text{diag}} &= 
 \frac12\partial_\lambda[\ln J_\lambda(\omega_0)] \left( \begin{array}{cc}
         e^{-2\xi} & 0 \\ [1.5ex] 0 &  e^{2\xi}
       \end{array}
      \right)  \\[0.2cm]
 \mathsf{D} &=  \left( \begin{array}{cc}
        \cos(\theta/2) & -\sin(\theta /2) \\ [1.5ex] \sin(\theta /2) &  \cos(\theta /2)
      \end{array}
     \right) \, .   \label{D}
\end{align}
Therefore, instead of considering the SLD a degree-2 polynomial in the canonical operators $\vec{Z}$ as in (\ref{SLDgaussian}),
one can always recast it as being quadratic in the rotated canonical operators $\vec{Z}'$, defined as
\begin{equation}
 \vec{Z}'=\mathsf{D^T}\vec{Z}= \left[ \begin{array}{c}
         X(\frac{\theta}{2}) \\ [1.5ex] X(\frac{\theta+\pi}{2}) \end{array}  \right]  \, , \label{rotatedop}
\end{equation}
obtaining:
\begin{equation}
\mathcal{L}_\lambda= \vec{Z}'^T \Phi_{\mathrm{diag}}\vec{Z}'=\mathrm{c}_{\frac{\theta}{2}} X^2\Big ( \frac{\theta}{2} \Big ) + 
 \mathrm{c}_{\frac{\theta+\pi}{2}} X^2\Big (\frac{\theta+\pi}{2}\Big ) \, ,
\end{equation}
where:
\begin{subequations}\begin{align}
 \mathrm{c}_{\frac{\theta}{2}} &=  \frac12\partial_\lambda[\ln J_\lambda(\omega_0)] e^{-2\xi}  \label{c1}  \\
 \mathrm{c}_{\frac{\theta+\pi}{2}}  &=  \frac12\partial_\lambda[\ln J_\lambda(\omega_0)] e^{2\xi} \, .  \label{c2}
\end{align}\end{subequations}
Now, considering that the estimation of the parameter $\lambda$ performs better when employing the probe initialized in a highly squeezed vacuum state, the ratio $ \mathrm{c}_{\frac{\theta}{2}}/\mathrm{c}_{\frac{\theta+\pi}{2}}= e^{-4\xi} $ favours the quadrature operator $X(\frac{\theta+\pi}{2}) $, so that the SLD can be well approximated by:
\begin{equation}
 \mathcal{L}_\lambda \simeq  \mathrm{c}_{\frac{\theta+\pi}{2}} X\Big (\frac{\theta+\pi}{2}\Big )^2 \, . \label{Lapprox}
\end{equation}
The important result of Eq.~(\ref{Lapprox}) shows that, under the weak-coupling expansion $\alpha\ll 1$ and for moderately high squeezing, the optimal observable on the probe is the quadrature which has been squeezed in the initial preparation, namely $\varphi_\text{opt}=\frac{\theta+\pi}{2}$, as one may expect, since squeezing, basically, means reducing the variance of a particular quadrature. Even though this is an approximated result, it gives a deep insight and a simple explanation on the general results found in the not approximated case (see, e.g., Fig.~\ref{f:FI_wc_phi}).

\section{Conclusions} \label{s:conclusions}
In this paper we have suggested a characterization scheme for structured environments
based on the use of continuous-variable quantum probes, focussing on the estimation of the
cutoff frequency of Ohmic environments. In particular, we have discussed 
in details how to optimize the extraction of information, i.e. how to maximize the QFI 
of  the cutoff frequency, depending on the initial Gaussian preparation of the probe 
and on the interaction between the probe and the environment. 
\par
A probe prepared in a squeezed vacuum state outperforms any other single-mode Gaussian-state initialization, in terms of the absolute value of the QFI, which grows for higher squeezing. 
Moreover, the QFI grows with the interaction time and temperature  
(see Fig.~\ref{f:QFI_wc_s_time}), but decreases for increasing thermal noise 
in the probe state (see Fig.~\ref{f:QFImax_nth_xi}). We also found that the non-Markovian character
of the interaction is not a resource for the present parameter estimation, as better estimation is obtained in the Markovian case, i.e. for long interaction times. Upon considering feasible
measurements, we found that optimality in homodyne detection, meant as the ratio between FI 
and QFI, is achieved for high values of initial squeezing, which increases the signal-to-noise ratio associated to the Fisher information (see Fig.~\ref{f:FIQFI0_FImax}). When thermal noise comes 
into play optimality is attained more quickly, at the expense of a lower SNR. Rather intuitively, 
we found that the optimal quadrature to be measured in order to maximize the FI, 
is that corresponding to the one being initially mostly squeezed (see Fig.~\ref{f:FI_wc_phi}). This result has been justified analytically in an approximated case in Sec.~\ref{s:ApprAlpha}.
\par
From a practical point of view, the estimation protocol here proposed suggests a tuning of 
the probe frequency to those we termed {\em sweet spots}, in order to maximize the precision in the estimation of the cutoff frequency. The positions of the sweet spots have been analyzed in terms of all the parameters of the evolved probe state (see Fig.~\ref{f:QFImax_nth_xi} and Fig.~\ref{f:QFI_wmax_nc}). Important results have been derived by a weak-coupling expansion 
(see Section \ref{s:ApprAlpha}), where the position of the sweet spots of 
both the QFI and the FI depends only on the spectral density class
(see Fig.~\ref{f:QFIapprox_wmax_s}). 
\par
In order to complete the analysis for an initial Gaussian state, we have also 
considered DSTS preparation of the probe and studied the coherent contribution 
of displacement. We found that the effect of a coherent amplitude in the probe state is to 
enhance the estimation performances, with a significant increase in the case of low-temperature environments (see Fig.~\ref{f:DSTS_H}). This boost comes together with a shift of the sweet spot position, which has been quantitatively studied, as reported in Fig.~\ref{f:QFI_wmax_nc}. When a continuous-variable probe state is considered, the total amount of energy of the state is the benchmark for a fair comparison of the probe performances. In the case of low-temperature environments, given a fixed fraction of thermal energy, when the total energy of the state is increased, the most convenient strategy, for parameter estimation, is to employ the remaining fraction into displacement rather than squeezing (see Fig.~\ref{f:fsOpt}).
\par
Our results have shown how squeezing may be employed as a resource to enhance estimation
of the cutoff frequency and pave the way for further developments as the probing of 
different structured environments and the comparison with quantum probes of different nature,
as discrete-variables or CV multimode ones.
\begin{acknowledgments}
This work has been supported by EU through the project QuProCS (Grant 
Agreement 641277). The authors thank D. Tamascelli, M.~G. Genoni, S. Olivares and 
C. Benedetti for useful discussions.
\end{acknowledgments}

\end{document}